\documentclass[aps,prd,english,superscriptaddress,11pt,notitlepage]{revtex4}
	\usepackage[colorlinks=true, a4paper=true, pdfstartview=FitV,
linkcolor=blue, citecolor=blue, urlcolor=blue]{hyperref}

\usepackage{amsmath}
\usepackage{amssymb}
\usepackage{graphicx}
\usepackage{hhline}
\usepackage{textcomp}
\makeatletter
\usepackage{babel}

\usepackage[active]{srcltx}

\newcommand{\be}{\begin{equation}}
\newcommand{\ee}{\end{equation}}
\newcommand{\bea}{\begin{eqnarray}}
\newcommand{\eea}{\end{eqnarray}}
\newcommand{\bean}{\begin{eqnarray*}}
\newcommand{\eean}{\end{eqnarray*}}
\font\upright=cmu10 scaled\magstep1
\font\sans=cmss12
\newcommand{\ssf}{\sans}
\newcommand{\stroke}{\vrule height8pt width0.4pt depth-0.1pt}
\newcommand{\Z}{\hbox{\upright\rlap{\ssf Z}\kern 2.7pt {\ssf Z}}}

\newcommand{\C}{{\rlap{\upright\rlap{C}\kern 3.8pt\stroke}\phantom{C}}}
\newcommand{\R}{\hbox{\upright\rlap{I}\kern 1.7pt R}}
\newcommand{\HH}{\hbox{\upright\rlap{I}\kern 1.7pt H}}
\newcommand{\CP}{\hbox{\C{\upright\rlap{I}\kern 1.5pt P}}}
\newcommand{\identity}{{\upright\rlap{1}\kern 2.0pt 1}}
\newcommand{\half}{\frac{1}{2}}
\newcommand{\quart}{\frac{1}{4}} 
\newcommand{\pr}{\partial}

\def\ir3{\int_{\mathbb{R}^{3}}}

\begin{document}

\title{Kink Moduli Spaces -- Collective Coordinates Reconsidered}

\author{N.~S. Manton}
\email[]{N.S.Manton@damtp.cam.ac.uk}
\affiliation{Department of Applied Mathematics and Theoretical Physics,
University of Cambridge,
Wilberforce Road, Cambridge CB3 0WA, U.K.}
\author{K. Ole\'{s}}
\email[]{katarzyna.slawinska@uj.edu.pl}
\affiliation{Institute of Physics,  Jagiellonian University, Lojasiewicza 11, Krak\'{o}w, Poland}
\author{T. Roma\'{n}czukiewicz}
\email[]{tomasz.romanczukiewicz@uj.edu.pl}
\affiliation{Institute of Physics,  Jagiellonian University, Lojasiewicza 11, Krak\'{o}w, Poland}
\author{A. Wereszczy\'{n}ski}
\email[]{andrzej.wereszczynski@uj.edu.pl}
\affiliation{Institute of Physics,  Jagiellonian University, Lojasiewicza 11, Krak\'{o}w, Poland}

\begin{abstract}
 Moduli spaces -- finite-dimensional, collective coordinate 
manifolds -- for kinks and 
antikinks in $\phi^4$ theory and sine-Gordon theory are reconsidered. The 
field theory Lagrangian restricted to moduli space defines a reduced 
Lagrangian, combining a potential with a kinetic term that can be 
interpreted as a Riemannian metric on moduli space. Moduli spaces
should be metrically complete, or have an infinite potential on their 
boundary. Examples are constructed for both kink-antikink and 
kink-antikink-kink configurations. 
The naive position coordinates of the kinks and antikinks sometimes need 
to be extended from real to imaginary values, although the field remains 
real. The previously discussed null-vector problem for 
the shape modes of $\phi^4$ kinks is resolved by a better coordinate 
choice. In sine-Gordon theory, moduli spaces can be constructed using 
exact solutions at the critical energy separating scattering and breather 
(or wobble) solutions; here, energy conservation relates 
the metric and potential. The reduced dynamics on these moduli 
spaces accurately reproduces properties of the exact solutions 
over a range of energies.
\end{abstract}

\maketitle

\vspace*{0.2cm}

\section{Introduction}

In field theory one often encounters particle-like 
solitons \cite{Ra1, DJ, MS}. We consider here the kink-solitons 
in Lorentz-invariant theories in one space dimension, and will 
focus on the well-known examples of $\phi^4$ and sine-Gordon 
kinks. Because they are topologically distinct from the vacuum 
field, these kinks are stable, and they can move at any speed less 
than the speed of light (1 in our units). The energy of a kink at 
rest is its mass, and each kink has an associated 
antikink with the same mass, obtained by spatial reflection. A
kink and antikink can coexist dynamically in the vacuum sector of 
the theory, but because they attract, there is no static 
kink-antikink solution.  

The fundamental distinction between field dynamics and particle
dynamics is that fields have infinitely many degrees of freedom, 
whereas particles have finitely many. The finitely many degrees of 
freedom of solitons are known as collective coordinates, or
moduli -- a concise nomenclature derived from pure mathematics 
that we use here in the context of kinks and antikinks. It remains an 
interesting challenge to construct a finite-dimensional 
approximation to the field dynamics -- a reduced dynamics on moduli 
space -- that adequately describes kink-antikink dynamics.
In this paper we aim for an improved understanding of these moduli
spaces and their geometry. Our models combine a Riemannian 
metric on moduli space with a potential energy function. Such models are
not new. Generally, the moduli space is a curved, finite-dimensional 
submanifold of the infinite-dimensional field 
configuration space, the space of all (static) field configurations at a 
given instant satisfying the boundary conditions for finite energy, and 
in the desired topological sector. The field theory has a kinetic
term, quadratic in time derivatives of the field and positive 
definite, that defines an (infinite-dimensional) metric. Restricting this to
time-dependent fields moving through moduli space (i.e., static field 
configurations with time-dependent moduli) gives a positive-definite 
expression quadratic in time derivatives of the moduli, defining 
a metric on moduli space. The field theory potential energy 
(which includes the field gradient term) restricts to a potential 
energy on moduli space. The reduced dynamics on moduli space is 
defined by the Lagrangian that combines this metric and 
potential. This is a natural Lagrangian system in the 
sense of Arnol'd \cite{Arn}. 

In naive models of particle dynamics, based on Newton's equations, the
kinetic terms define a Euclidean metric on the space of positions, and
the potential alone is responsible for  generating forces and
particle scattering. By contrast,
for certain types of soliton, usually in more than one space
dimension, there are no static forces between solitons; these are the
solitons of Bogomolny (or BPS) type \cite{Bo}. Examples are abelian Higgs
vortices at critical coupling \cite{JT} and non-abelian 
Yang--Mills--Higgs monopoles \cite{Ma8}. Here, there is a
canonically-defined moduli space of static $N$-soliton solutions 
(solitons and antisolitons never occur simultaneously), on which there
is a Riemannian metric defined by the field theory kinetic terms. 
The potential energy is a constant proportional to 
the topological charge, having no effect, so the 
dynamics on moduli space depends on the metric alone and 
the motion is along geodesics at constant speed 
\cite{Ma1, AH, Sam}. This is usually a good approximation to 
the soliton dynamics in the field theory, provided the 
motion is not highly relativistic. The geodesic motion can be
nontrivial, and result in soliton scattering over a
wide range of angles. There are also examples of solitons close to those
of Bogomolny type, for example, vortices close to critical
coupling. The potential is then non-zero but small, and modifies 
the geodesic motion on moduli space. 
In the examples of kink-antikink dynamics that we study here, the
metric on the moduli space is not Euclidean, and the potential
energy is not small. The metric and potential play
equally important roles.

For kinks and antikinks, there is no canonically-defined 
moduli space. However, basic principles can be learned from the Bogomolny 
examples. First, the moduli space should be a metrically complete 
submanifold of field configuration space, or the potential should be
infinite on any boundary. If not, a dynamical trajectory can reach the 
boundary in finite time, and it is unclear what happens next. Second, the
moduli space should be smoothly embedded in the field configuration
space, otherwise the moduli space motion cannot smoothly approximate the
true field dynamics. Apparent singularities in the moduli space metric 
sometimes occur, but these can be resolved by a better choice of 
coordinates, as we will show. 

In Section 2 we introduce the notion of moduli space 
dynamics in $\phi^4$ theory, and illustrate it with the simple example 
of single kink motion. A static kink interpolates between the two vacuum 
field values $\pm 1$, and has the simple form 
$\phi(x) = \tanh(x-a)$, where $a$ is the modulus representing the 
position of the kink. The antikink with position 
$a$ is $\phi(x) = -\tanh(x-a)$, and is the reflection of the kink. 
In the moduli space dynamics, $a$ becomes a function of time $t$. 
We defer discussion of kink-antikink dynamics in $\phi^4$ theory 
until Section 4, because it is relatively complicated. 

Instead, in Section 3 we consider sine-Gordon (sG) theory, and
construct a moduli space model for kink-antikink dynamics 
there. As sG theory is exactly integrable, we can compare the moduli space
dynamics with the known exact solutions, for both kink-antikink 
scattering and the lower-energy sG breathers,
which can be interpreted as bounded kink-antikink motion.
We show that having a non-trivial metric on the moduli space is 
important. A model with
only a potential energy depending on the kink-antikink separation
is less successful. We also investigate kink-antikink-kink dynamics,
where again an exact solution is known. One feature of sG theory
is that for each kink and antikink we need just one
collective coordinate. Our models do not identically reproduce the
behaviour of the exact solutions because the field
configurations we use do not vary as the soliton speeds change -- the
solitons do not Lorentz contract. The modelling is therefore 
non-relativistic in character, and yet it can cope quite
well with kinks and antikinks close to annihilation, a
fundamentally relativistic process, where the initial 
rest energy of the solitons converts entirely into kinetic energy.   

In Section 4 we return to the kinks and antikinks in non-integrable 
$\phi^4$ field theory. There are field 
configurations with a string of well-separated, alternating kinks 
and antikinks along the spatial line, whose field value varies 
between close to $-1$ and $+1$. A kink and antikink are in some sense 
identical particles, because of this forced alternation. 
There have been numerous studies of kink-antikink dynamics 
in $\phi^4$ theory, starting with the pioneering work of 
\cite{Sug, Mosh, CSW} and others; for a review, see
\cite{KG}. Because the field dynamics is complicated, it is difficult 
to devise an approximate dynamics with a finite number of moduli,
modelling the kinks and antikinks as particles.

The simplest model for a kink and antikink would have just two degrees 
of freedom, their positions. However, it is well known 
that in $\phi^4$ theory, a kink has an internal vibrational degree of 
freedom -- a shape mode -- because the linearised field equation in the 
kink background has one localised field vibration mode whose 
frequency is below the continuum of radiation modes that can disperse to 
spatial infinity. More sophisticated models therefore allow the kink and 
antikink to have two degrees of freedom each, allowing for transfer of 
energy between the positional dynamics and the shape modes.
Moduli space dynamics conserves energy and doesn't account for the 
conversion of energy into radiation. Nevertheless, it can clarify 
the mechanism by which energy is transferred in and 
out of the shape modes, and by coupling the moduli to other field 
modes, the mechanism and timing of the production of radiation may be
better understood.

It is recognised that some finite-dimensional models of $\phi^4$ 
kink-antikink dynamics have had problems. The manifold on which the 
finite-dimensional dynamics takes place has not always been complete, and 
if complete then unsuitable coordinates have sometimes been chosen. 
Problems tend to occur when the kink and antikink are close 
together and about to annihilate. We will show here that these problems 
can be resolved.
It might be thought that the problems are intrinsic -- that the notion of 
kink and antikink as separate objects is bound to fail when they are 
about to annihilate. However this is not the case. Field simulations
in $\phi^4$ theory show that in a collision of a kink and 
antikink, after rather dramatic behaviour while they are close together, 
they can emerge relatively unscathed and separate 
to infinity with limited transfer of energy to radiation. Even if
they completely annhilate into radiation, this can be a relatively
slow process.

There are some rather simple formulae for kink-antikink field configurations 
in $\phi^4$ theory that have been proposed previously, and we will use 
these together with some variants. The basic formula is simply to 
add (superpose) the exact static kink solution to the exact antikink. A
field shift by a constant is needed to satisfy the 
boundary conditions. We will also consider the extension of this formula, 
due to Sugiyama \cite{Sug}, where the shape modes for the kink and the 
antikink are included, with arbitrary amplitudes. We resolve the problem of 
the null-vector that can occur here, noted by Takyi and Weigel \cite{TW}. 
A further, novel variant introduces a weight factor multiplying the usual 
kink and antikink profiles. The aim is not to add further moduli, but 
to more accurately model the field dynamics when 
the kink and antikink are close to annihilating.

We will also consider, in Section 5, the configurations 
of $\phi^4$ kinks and antikinks that 
occur in the recently proposed iterated kink equation \cite{MOW}. The 
$n$th iterate $\phi_n$ provides a moduli space of dimension $n$ 
for a total of $n$ kinks and antikinks. Iterated kinks do not 
allow for the usual shape mode deformations, but do allow for 
arbitrary kink positions, and smoothly allow a kink and antikink 
to pass through the vacuum configuration. The 
kink-antikink configurations $\phi_2$ are weighted 
superpositions of the usual kink and antikink, and this is
part of the motivation for introducing weight factors.

In Section 6 we study spatially-antisymmetric kink-antikink-kink 
configurations in $\phi^4$ theory, using a naive superposition of profiles, 
a weighted superposition of profiles, and solutions $\phi_3$ in the 
iterated kink scheme. These field configurations differ, although they agree 
when the two kinks are well separated from the central 
antikink. We quantitatively compare these three descriptions.

One remarkable discovery, seen here in several guises, is that the 
moduli space dynamics in the kink-antikink and kink-antikink-kink 
sectors must sometimes be interpreted as particle motion where 
the particle positions scatter from real to imaginary values. 
(In detail, this depends on whether there are weight factors or not.) 
This is a curious extension of the well-known 90-degree scattering in
head-on collisions of higher-dimensional topological solitons,
including vortices \cite{SR, Sam}, monopoles \cite{AH} and Skyrmions 
\cite{AM2}. Two such solitons approaching each
other symmetrically along the $x$-axis, say, collide smoothly at 
the origin and emerge back-to-back along the $y$-axis. For 
the 3-dimensional solitons, the scattering plane is determined 
by the initial conditions, e.g., the relative orientation of Skyrmions. 
In a $\phi^4$ kink-antikink collision, the incoming positions are 
$-a$ and $a$, with $a$ approaching zero. After the collision, the
positions are $i{\tilde a}$ and $-i{\tilde a}$, with ${\tilde a}$ 
real and increasing from zero. The field remains real. As ${\tilde a}$
increases further, the potential energy becomes large, and at some 
point the motion stops and reverses. The relevant moduli space is not
complete unless one allows $a$ to become imaginary, and this is 
because the good coordinate is really $a^2$, which can become 
negative. 90-degree scattering is possible in the complex plane 
of the position coordinate $a$ because of the 
identity of the kinks, and the symmetry of the field configuration 
under interchange of $a$ and $-a$. 

\section{Moduli Spaces in $\phi^4$ Theory}

The manifold and metric structure of the infinite-dimensional 
$\phi^4$ field configuration space is quite simple. We recall it 
here and describe how it can be used to
construct the metric on a finite-dimensional moduli space of fields. We
then apply this formalism to the basic example of the moduli space of
a single kink. This is an opportunity to fix conventions and notation before
tackling kink-antikink and kink-antikink-kink fields.

The Lagrangian of $\phi^4$ theory in one spatial dimension is
\be
L = \int_{-\infty}^{\infty} \left(\half{\dot \phi}^2 - \half \phi'^2 -
\half (1 - \phi^2)^2 \right) \, dx \,,
\ee
where $\phi(x,t)$ is a real field taking arbitrary values. An overdot 
denotes a time derivative, and a prime denotes a spatial
derivative. The Euler--Lagrange field equation is the 
nonlinear Klein--Gordon equation
\be
\ddot \phi - \phi'' - 2(1 - \phi^2)\phi = 0 \,. \label{eom}
\ee
The Lagrangian can be split into positive definite kinetic and potential
terms as $L = T - V$, where $T$ just involves the integral of
${\dot\phi}^2$, and the remaining terms contribute to $V$. The splitting
of kinetic from gradient terms hides the theory's Lorentz invariance,
but this seems to be necessary to obtain a moduli space dynamics for
kinks.

The theory has two vacua, $\phi = \pm 1$, and we restrict attention to
finite-energy field configurations that approach vacuum field values at 
spatial infinity. There are four topologically distinct sectors. One is the
vacuum sector where $\phi \to -1$ as $x \to \pm \infty$; the vacuum
itself is $\phi(x) = -1$ for all $x$. Another is the kink sector, 
where $\phi \to \mp 1$ as $x \to \mp\infty$. There is a second vacuum 
sector and an antikink sector which are similar.

Field configuration space is an infinite-dimensional affine
space, a space modelled on a linear space but having no
canonical origin. In the vacuum sector, one can express any field
configuration as $\phi(x) = -1 + \chi(x)$ where $\chi \to 0$ as $x \to
\pm\infty$. The shifted fields $\chi$ can be linearly combined, which
explains the affine structure.

For static fields, the field equation (\ref{eom}) has the first integral
\be
\phi' = 1 - \phi^2 \,.
\ee
The constant of integration is chosen so that a solution can
satisfy the vacuum boundary conditions $\phi \to \pm 1$ at spatial infinity,
and the sign choice for the square root ensures the solutions are
kinks rather than antikinks. The kink solutions are
\be
\phi(x;a) = \tanh(x-a)
\ee
where $a$, the further constant of integration, is arbitrary. 
$a$ represents the position of the kink centre, where $\phi = 0$. $a$ is
the modulus of the solution, and the moduli space of kink solutions is
the real line. The kink sector of field configuration space is again an affine
space, because the generic configuration in this sector can be written
as $\phi(x) = \tanh(x) + \chi(x)$, where $\chi$ has the same decay
properties as before, so different field configurations $\chi$ can be
linearly combined. We have chosen the kink centred at the origin as 
the origin of the affine space, but this is just a choice.

Each sector of field configuration space has a natural, 
infinite-dimensional Euclidean metric. The squared
distance between configurations $\phi_1$ and $\phi_2$ is
\be
s^2 = \int_{-\infty}^{\infty} (\phi_2(x) - \phi_1(x))^2 \, dx \,.
\label{s2}
\ee
The integrand is also the squared difference of $\chi$ fields,
which shows that the metrics in the vacuum and kink sectors
are essentially the same. Note the need for spatial integration. 
By considering field configurations with infinitesimal separation, 
$\phi$ and $\phi + \delta\phi$, we obtain the (Euclidean) Riemannian metric
on configuration space
\be
\delta s^2 = \int_{-\infty}^{\infty} (\delta\phi(x))^2 \, dx \,.
\label{ds2}
\ee

Any moduli space is a finite-dimensional submanifold of
field configuration space, where the configurations $\phi(x;{\bf y})$
depend on finitely many moduli (collective coordinates) $y^i$, denoted 
jointly as ${\bf y}$. Varying the moduli, we have the field variation
\be
\delta\phi(x;{\bf y}) = \frac{\pr \phi}{\pr y^i}(x;{\bf y}) \,\delta y^i \,.
\ee
The Riemannian metric restricted to the moduli space can therefore be
written as
$\delta s^2 = g_{ij}({\bf y}) \, \delta y^i \, \delta y^j$, where, from
eq.(\ref{ds2}), we see that
\be
g_{ij}({\bf y}) =  \int_{-\infty}^{\infty} 
\frac{\pr \phi}{\pr y^i}(x;{\bf y})
\frac{\pr \phi}{\pr y^j}(x;{\bf y}) \, dx \,. 
\label{metric-def}
\ee
This key formula is used to define the metric on moduli spaces
of kinks and antikinks, and allows one to study if the moduli space 
is metrically complete, and whether the coordinates have been well
chosen. Generally, the moduli space is curved if it is more than
1-dimensional. From (\ref{metric-def}) we infer that the kinetic energy for
a field evolution $\phi(x;{\bf y}(t))$ restricted to the moduli space is
\be
T = \frac{1}{2} g_{ij}({\bf y}) {\dot y}^i {\dot y}^j \,,
\ee
as $\dot\phi = \frac{\pr \phi}{\pr y^i} {\dot y}^i$.

For the 1-dimensional moduli space of a single kink, 
$\phi(x;a) = \tanh(x-a)$, the metric is 
$\delta s^2 = g(a) \,\delta a^2$ where
\be
g(a) = \int_{-\infty}^{\infty} \left(\frac{\pr \phi}{\pr a}(x;a)\right)^2
\, dx \,.
\ee
The derivative with respect to $a$ can be
traded for (minus) the derivative with respect to $x$, so
\be
g(a) = \int_{-\infty}^{\infty} \phi'^2(x;a) \, dx \,,
\ee
which is independent of the position $a$ of the kink.
This integral equals the mass $M$ (the total potential energy $V$) of 
the kink, a result that generalises to any type of scalar kink in a 
Lorentz-invariant theory in one dimension. This is because
\be
V = \int_{-\infty}^{\infty} \left( \half\phi'^2 
+ \half(1 - \phi^2)^2 \right) \, dx 
\ee
is minimised in its topological sector by the kink, and
in particular is minimised under a spatial rescaling of the kink profile. The
latter property implies that the two integrals contributing to $V$
are equal, and therefore $V$ equals the integral of $\phi'^2$. For
the $\phi^4$ kink, $V = M = \frac{4}{3}$ and
therefore $g(a) = \frac{4}{3}$.

The kinetic energy of a single kink, having a time-dependent modulus
$a(t)$, is therefore $\half M \dot a^2 = \frac{2}{3} {\dot a}^2$, 
the standard non-relativistic kinetic energy of a moving particle. As the 
potential energy is a constant, the reduced Lagrangian 
is
\be
L_{\rm red} = \frac{2}{3} {\dot a}^2 \,.
\ee
The equation of motion is ${\ddot a} = 0$, and the kink moves with 
constant velocity. There is no Lorentz contraction of the kink, and 
its speed is not constrained to be less than the speed of light.

Metrically, the single kink moduli space is the real line with its 
Euclidean metric scaled by a constant factor. $a$ is a good
coordinate, because the metric $g(a) = M$ is everywhere positive. We
will see examples later where a moduli space metric appears not
to be positive definite, but becomes so with a better choice of
coordinates.

\section{Sine-Gordon Theory}

A laboratory where we can test our ideas is the sine-Gordon (sG) field 
theory \cite{PSk} where the exact multi-soliton
solutions are known. The sG Lagrangian is
\be
L = \int_{-\infty}^{\infty} \left( \half {\dot \phi}^2 - \half {\phi'}^2 - (1
- \cos\phi) \right) dx \,.
\label{sGLagran}
\ee
Its field equation,
\be
{\ddot \phi} - \phi'' + \sin\phi = 0 \,,
\ee
has static kink (soliton) solutions
\be
\phi(x;a) = 4\arctan(e^{x-a})
\ee
interpolating between $0$ and $2\pi$. The antikink is 
$\phi(x;a) = -4\arctan(e^{x-a})$, interpolating between $0$ and 
$-2\pi$. Both the kink and antikink have mass $M = 8$.

In the moduli space dynamics, a single moving 
kink is modelled by
\be
\phi(x,t) = 4\arctan(e^{x-a(t)})
\ee 
and the reduced Lagrangian is $L_{\rm red} 
= \half M {\dot a}^2 = 4 {\dot a}^2$. 
Like a $\phi^4$ kink, an sG kink moves at constant velocity. 

\subsection{Kink-antikink dynamics in sine-Gordon theory}
\begin{figure}
\center \includegraphics[width=0.82\textwidth]{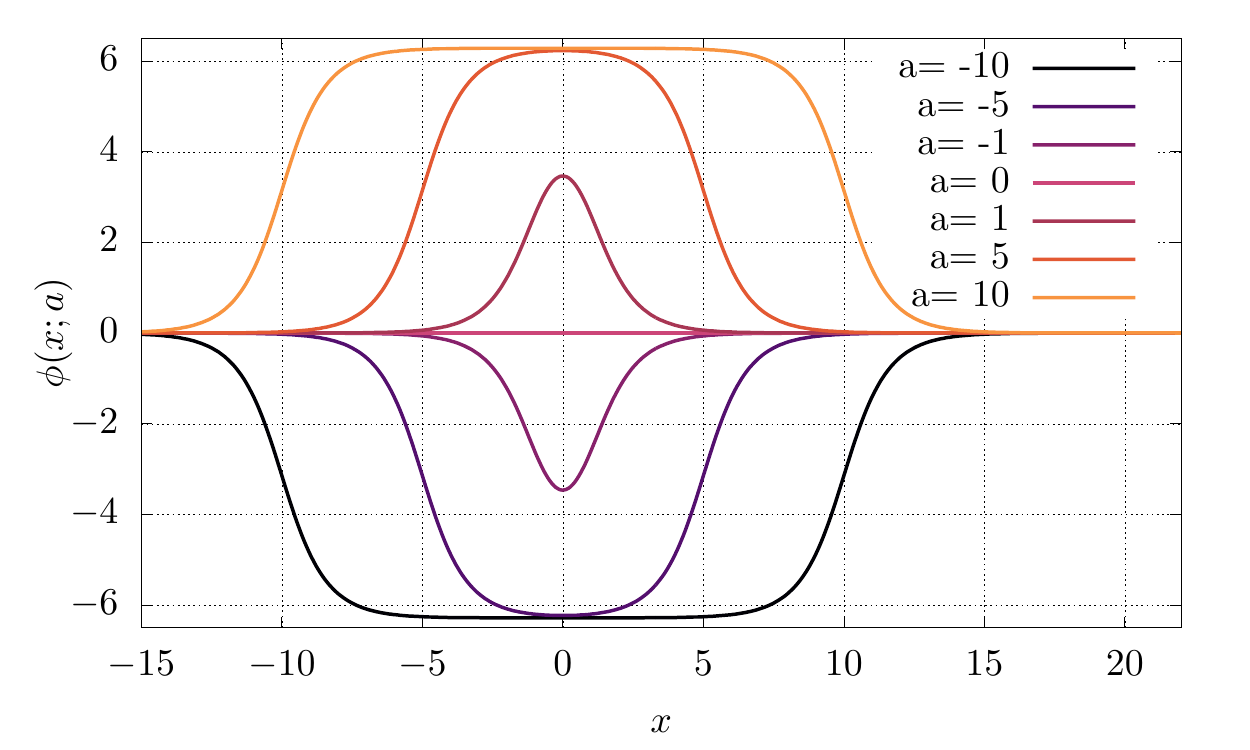}
\caption{Examples of the naive superposition of the sG kink and 
antikink (\ref{sGKK*}).}
\label{sG-2-naive} 
\end{figure}

The naive superposition of a kink centred at $-a$ and an antikink 
centred at $a$ is
\be
\phi(x;a) = 4 \arctan(e^{x+a}) - 4 \arctan(e^{x-a}) \,.
\ee
Applying the tangent subtraction formula to $\quart\phi$, this becomes
\be
\phi(x;a) = 4\arctan \left( \frac{\sinh(a)}{\cosh(x)} \right) \,,
\label{sGKK*}
\ee
where $a$ takes any real value. This is our moduli space of
kink-antikink sG field configurations. For $a \gg 0$ there
is a kink on the left and antikink on the right, and $\phi$ is
positive; for $a \ll 0$ the kink and antikink are exchanged
and $\phi$ is negative. When $a=0$, the configuration is the vacuum 
$\phi = 0$. Configurations for various values of $a$ are shown 
in Fig. \ref{sG-2-naive}.

To model kink-antikink dynamics (both scattering and periodic 
solutions, i.e., breathers) we treat $a(t)$ as a time-dependent modulus,
substitute \eqref{sGKK*} into the field theory Lagrangian 
\eqref{sGLagran} and integrate, thereby obtaining a reduced Lagrangian 
on moduli space of the general form
\be
L_{\rm red} =\frac{1}{2}g(a) \, \dot{a}^2 - V(a) \,. 
\label{sGeff}
\ee
The kinetic term of the field theory gives the moduli space 
metric
\be
g(a)= 16 \left( 1+ \frac{2a}{\sinh(2a)} \right)
\ee
and the remaining terms give the moduli space potential
\be
V(a)=16 \left[ 1 -  \frac{1}{2\cosh^2(a)} 
\left(1 + \frac{2a}{\sinh(2a)} \right) \right] \,. 
\ee
Note that the metric and potential satisfy 
\be
\frac{1}{2\cosh^2(a)} \, g(a) + V(a) = 16 \,,
\label{metricpotrel}
\ee
a relation that we will explain below. For large $a$, $g(a) \sim 16$ 
and $V(a) \sim 16 - 32e^{-2a}$, from which one can derive the 
well-known static force between an sG kink and antikink at 
large separation $2a$.

The time-dependence of the modulus $a$ can be found starting from 
the first integral of the equation of motion derived from \eqref{sGeff},
\be
\frac{1}{2}g(a) \, \dot{a}^2 + V(a) = E \,,
\label{sGenergy}
\ee
and then integrating once more. Both $V$ and the conserved energy $E$ 
include the rest masses of the kink and antikink, so the critical 
energy separating scattering solutions from breathers is $E = 16$. 
The motion on moduli space is implicitly given by
\be
\pm(t-t_0) = \int \sqrt{\frac{g(a)}{2(E-V(a))}} \, da \,,
\label{a(t)soln}
\ee
and can be calculated numerically.

Before presenting the solutions, let us recall the analogous exact solutions
of sine-Gordon theory. The breather solution
\be
\phi(x,t) = -4\arctan \left( \frac{\sqrt{1-\omega^2}\sin(\omega t)}
{\omega \cosh(\sqrt{1-\omega^2} \, x)} \right)
\label{breather}
\ee
has frequency $\omega$ in the range $(0,1)$ and energy
\be
E = 16\sqrt{1-\omega^2} \,.
\label{Eomega}
\ee
This solution exactly matches (\ref{sGKK*}) if we ignore the shape-changing 
factor $\sqrt{1-\omega^2}$ in $\cosh(\sqrt{1-\omega^2} \, x)$ and 
say that the breather has modulus dynamics given by 
\be
\sinh(a(t)) = -\frac{\sqrt{1-\omega^2} \sin(\omega t)}{\omega} \,.
\label{a(t)breath}
\ee

The kink-antikink scattering solution is the analytic
continuation of the breather, when the frequency becomes
imaginary. Setting $\omega = iq$ in \eqref{breather}, with $q$ 
real, we obtain the solution
\be
\phi(x,t) = -4\arctan \left( \frac{\sqrt{1+q^2} \sinh(q t)}
{q \cosh(\sqrt{1+q^2} \, x)} \right) \,.
\label{sGscattq}
\ee 
The reparametrisation $v = q/\sqrt{1+q^2}$ leads to the more standard form
of the scattering solution
\be
\phi(x,t) = -4\arctan \left( \frac{\sinh(\gamma vt)}
{v \cosh(\gamma x)} \right) \,,
\label{sGscattv}
\ee
where $\gamma = 1/\sqrt{1-v^2}$ is the usual Lorentz factor and
$\gamma v = q$. This describes a kink and 
antikink that approach each other with velocities $v$ and $-v$, 
having total energy $E = 16\gamma = 16\sqrt{1+q^2}$. By ignoring the 
Lorentz contraction factor $\gamma$ in $\cosh(\gamma x)$, we identify
the exact scattering solution as having modulus dynamics
\be
\sinh(a(t)) = -\frac{\sinh(\gamma vt)}{v} \,.
\label{a(t)scatt}
\ee 
The asymptotic form of the solution, for large negative 
$t$ and $x$, is the incoming kink
\be
\phi(x,t) \sim 4\arctan (e^{\gamma(x-vt) - \log v}) \,,
\ee
and the outgoing kink (with field value shifted down by $2\pi$) has 
the opposite positional shift. 

Particularly relevant is the exact sG solution 
with the critical energy $E=16$, 
\be
\phi(x,t) = -4\arctan \left( \frac{t}{\cosh(x)} \right) \,,
\label{critbreather}
\ee
which can be regarded either as a scattering solution where the initial
incoming velocities have decreased to zero, or as a breather of infinite
period, where the kink and antikink reach spatial infinity with 
zero velocity. It evolves precisely through the configurations in our 
moduli space, with
\be
\sinh(a(t)) = -t \,.
\label{critE}
\ee
Equation \eqref{critE} also solves the equation of motion
for $a(t)$ on the moduli space, because a solution
of the field equation, being a stationary point of the action for 
unconstrained field variations, is automatically a 
stationary point of the action for a smoothly embedded set of 
constrained fields (the fields in the moduli space), provided 
the Lagrangian of the constrained problem is the restriction
of the Lagrangian of the unconstrained problem, as here.
This has an interesting consequence. Differentiating \eqref{critE}, 
we see that $\cosh(a(t)) \, {\dot a} = -1$, and substituting this into 
the energy conservation equation \eqref{sGenergy} we derive the relation
\eqref{metricpotrel} between the metric and potential on moduli
space.

\begin{figure}
\includegraphics[width=0.92\textwidth]{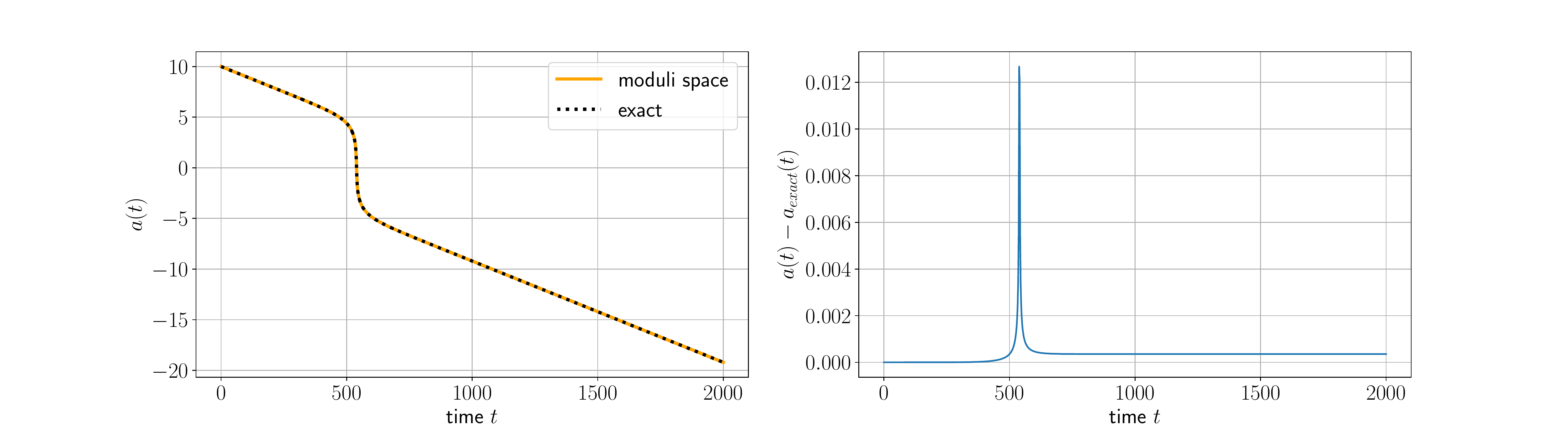}
\includegraphics[width=0.92\textwidth]{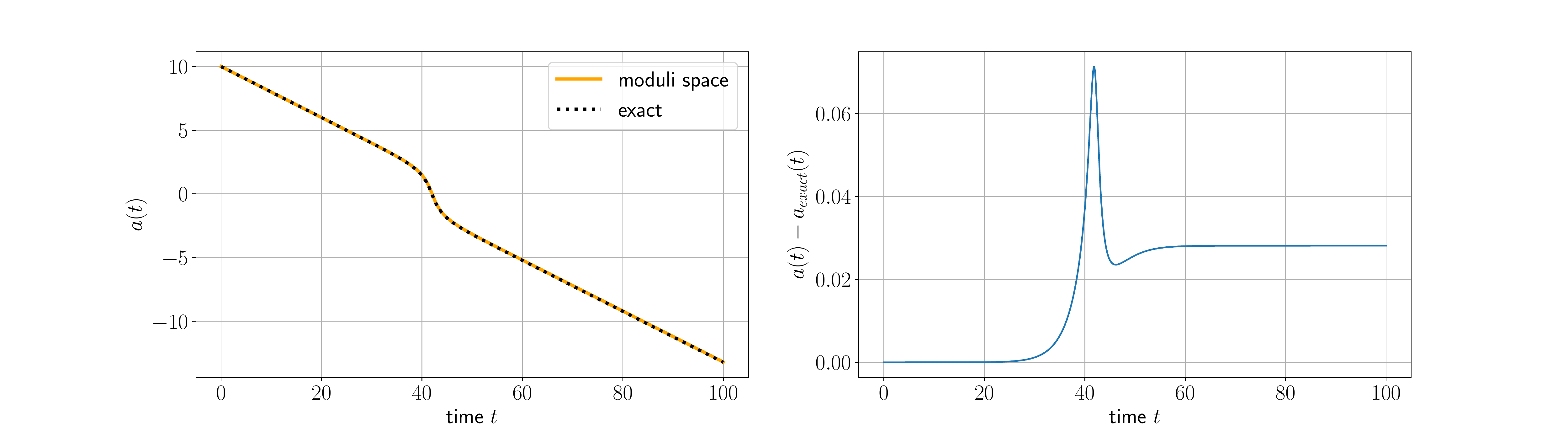}
\caption{Comparison of the dynamics of the modulus $a$ derived from
the moduli space evolution (\ref{a(t)soln}), with the exact
nonrelativistic solution (\ref{nonrelexact}). The small difference is
shown magnified on the right. Here the initial kink
velocity is $v=0.01$ (upper panels) and $v=0.2$ (lower panels).}
\label{profiles}
\end{figure}
\begin{figure}
\center \includegraphics[width=0.72\textwidth]{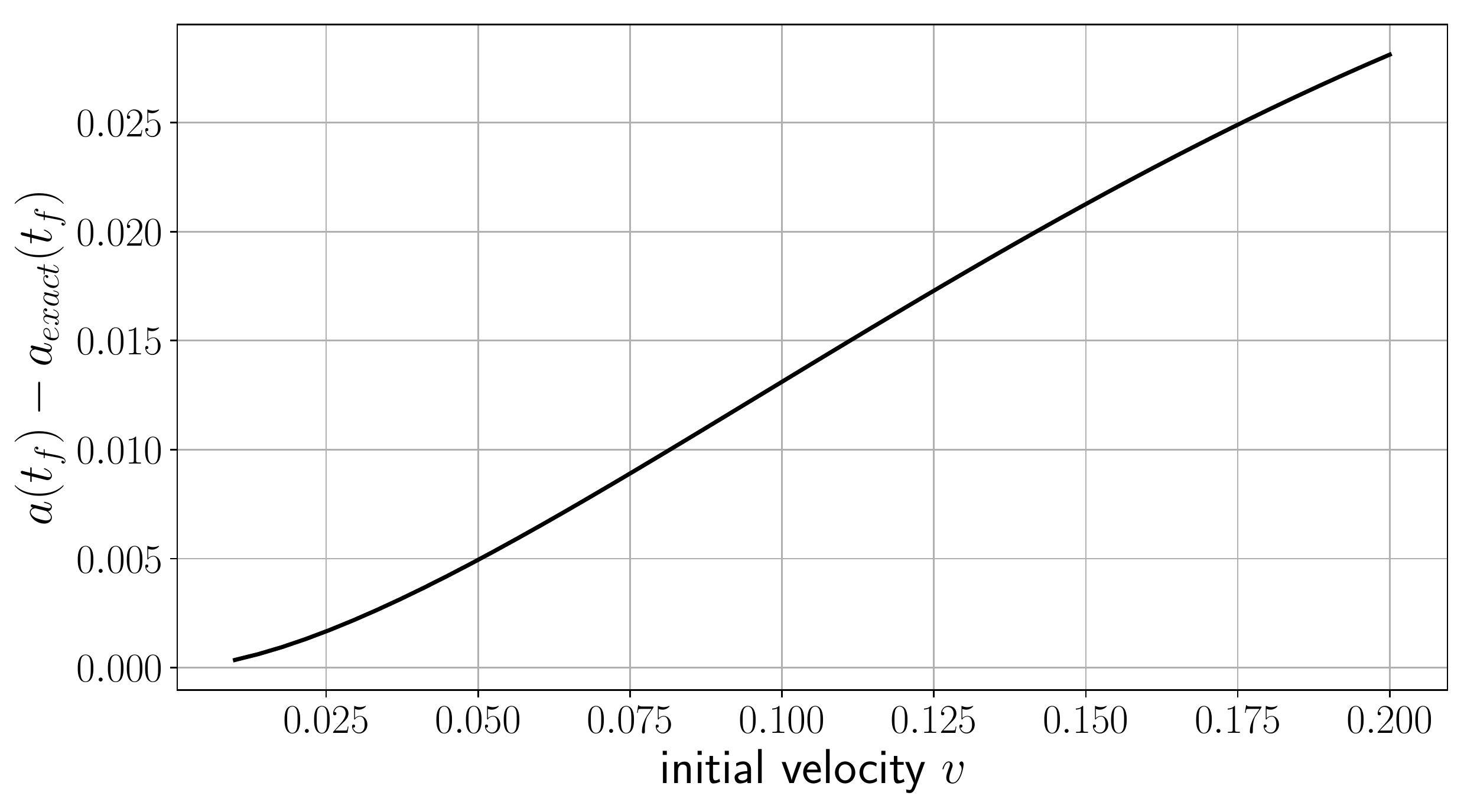}
\caption{Difference between the asymptotic value of the modulus $a$
derived from the moduli space evolution (\ref{a(t)soln}), and from the
exact nonrelativistic solution (\ref{nonrelexact}), as a function of 
velocity. Here $t_f=100$.}
\label{deviation}
\end{figure}
\begin{figure}
\center \includegraphics[width=0.72\textwidth]{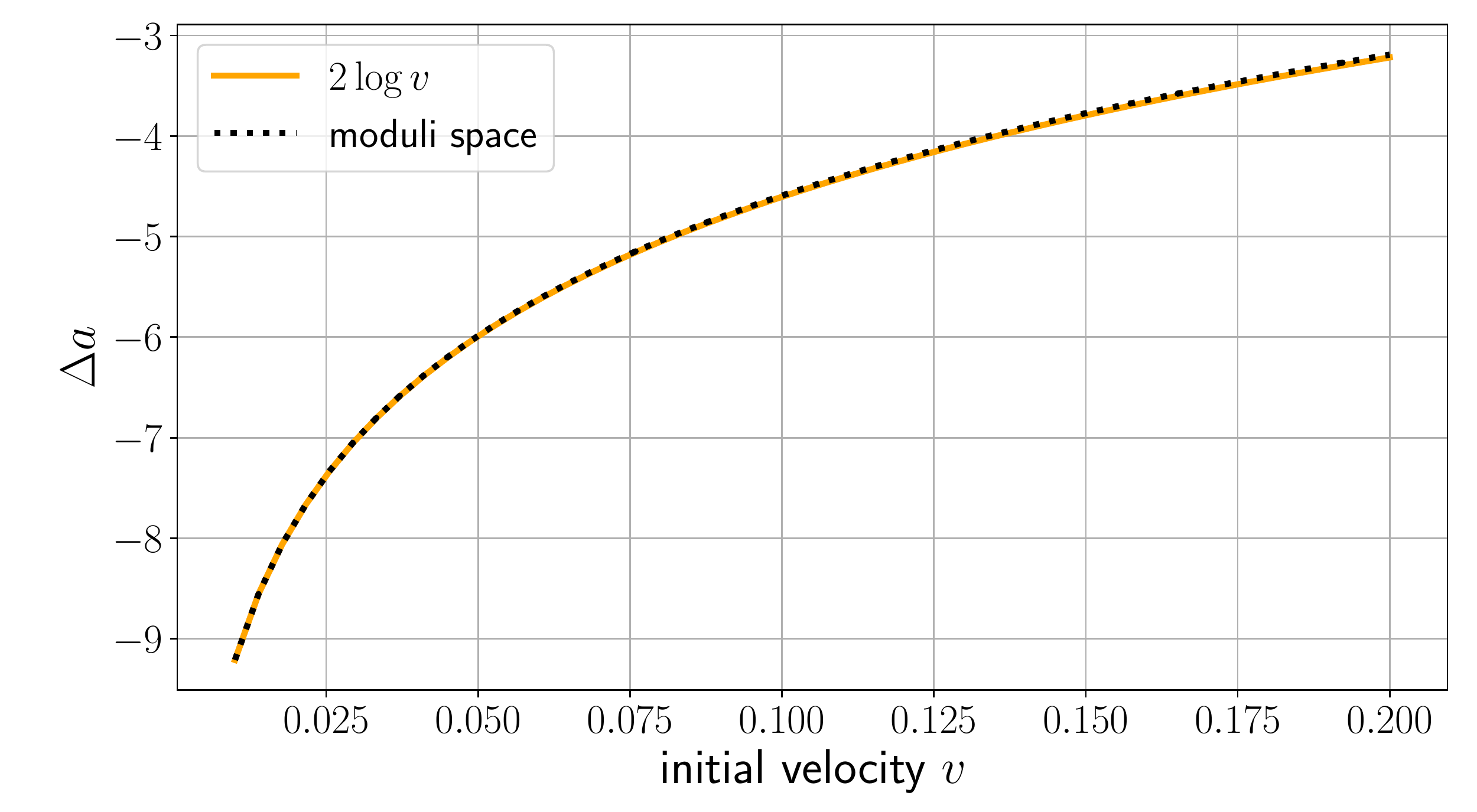}
\caption{The positional shift $\Delta a$ in a kink-antikink
collision derived from the moduli space evolution 
(\ref{a(t)soln}), compared with the exact nonrelativistic result
$2\log v$.}
\label{shift}
\end{figure}

Let us now investigate the accuracy of the moduli space dynamics above
and below the critical energy $E=16$. If $E > 16$ then $a \to \mp \infty$ as 
$t \to \pm \infty$ and there is kink-antikink scattering. The kink 
effectively passes through the antikink. In Fig. \ref{profiles} we compare the 
moduli space evolution of $a(t)$ according to 
\eqref{a(t)soln} (solid line) with the nonrelativistic version of
the exact evolution \eqref{a(t)scatt} (dotted line),
\be
\sinh(a(t)) = -\frac{\sinh(v t)}{v} \,. 
\label{nonrelexact}
\ee
The agreement is very good. There is a very small discrepancy that
grows with the velocity, shown in Fig. \ref{deviation}. In Fig. 
\ref{shift} we show the positional shift of $a$ away from a linear
evolution in time, due to the collision. The solid line represents 
the exact, nonrelativistic shift
\be
\Delta a = 2\log v \,,
\ee
while the dotted line is our numerical result from the moduli 
space dynamics. The agreement is again striking. 

If $E<16$ then $a$ oscillates in the moduli space dynamics between 
turning points $\pm a_{\rm max}$ (where the kinetic energy vanishes) given by
\be
V(\pm a_{\rm max}) = 16\left[ 1- \frac{1}{2\cosh^2(a_{\rm max})}
\left(1+\frac{2a_{\rm max}}{\sinh(2a_{\rm max})}\right) \right] = E \,.
\ee
This motion corresponds to a breather. Its period, as a function 
of $a_{\rm max}$, is
\be
\sqrt{2} \int_{-a_{\rm max}}^{a_{\rm max}} 
\sqrt{\frac{g(a)}{V(a_{\rm max})-V(a)}} \, da \,, 
\ee
\begin{figure}
\center \includegraphics[width=0.92\textwidth]{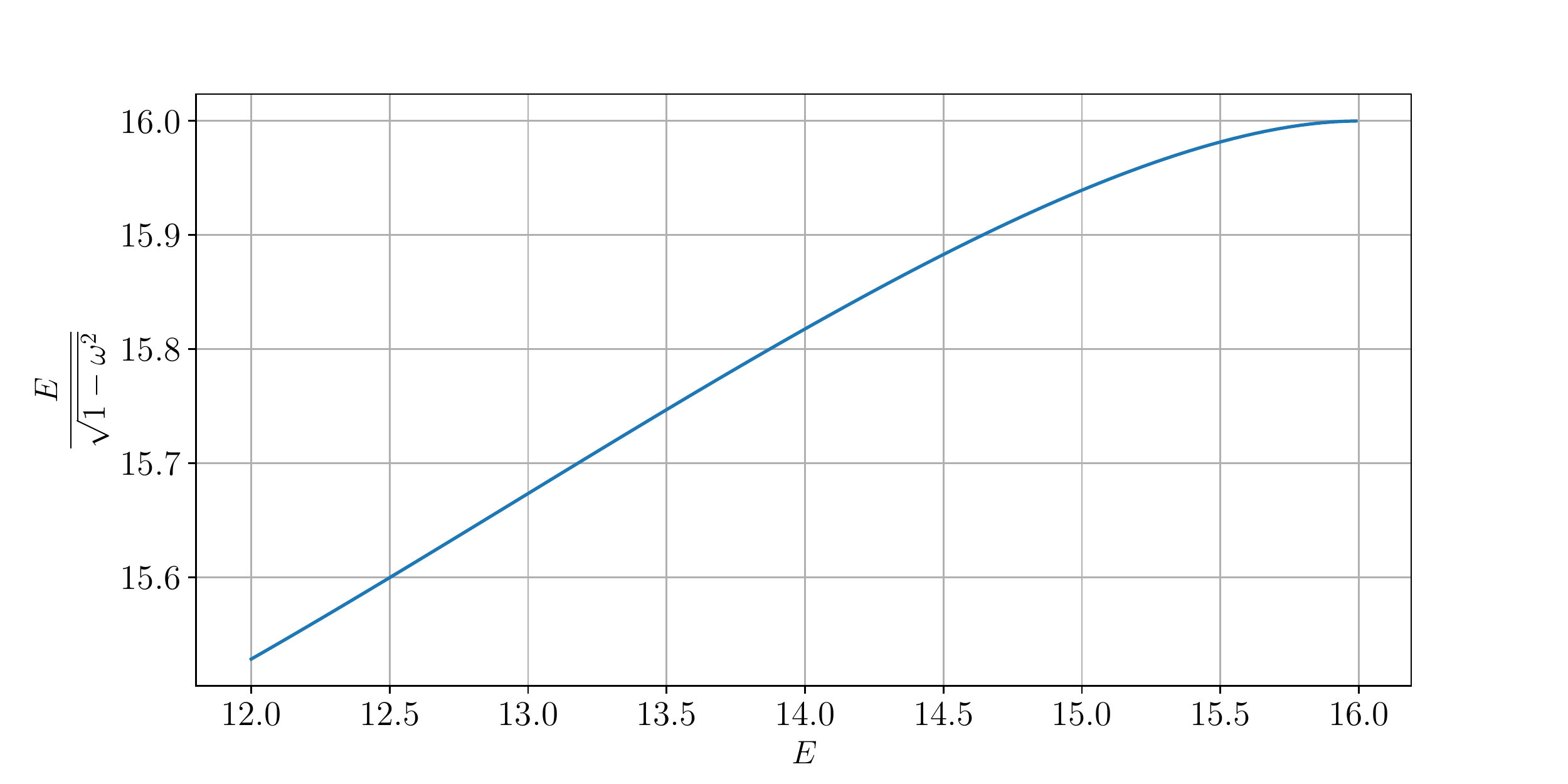}
\caption{ $E/\sqrt{1-\omega^2}$  vs. $E$ for the periodic solutions 
of the moduli space dynamics.}
\label{energy-freq}
\end{figure}
\begin{figure}
\center \includegraphics[width=0.82\textwidth]{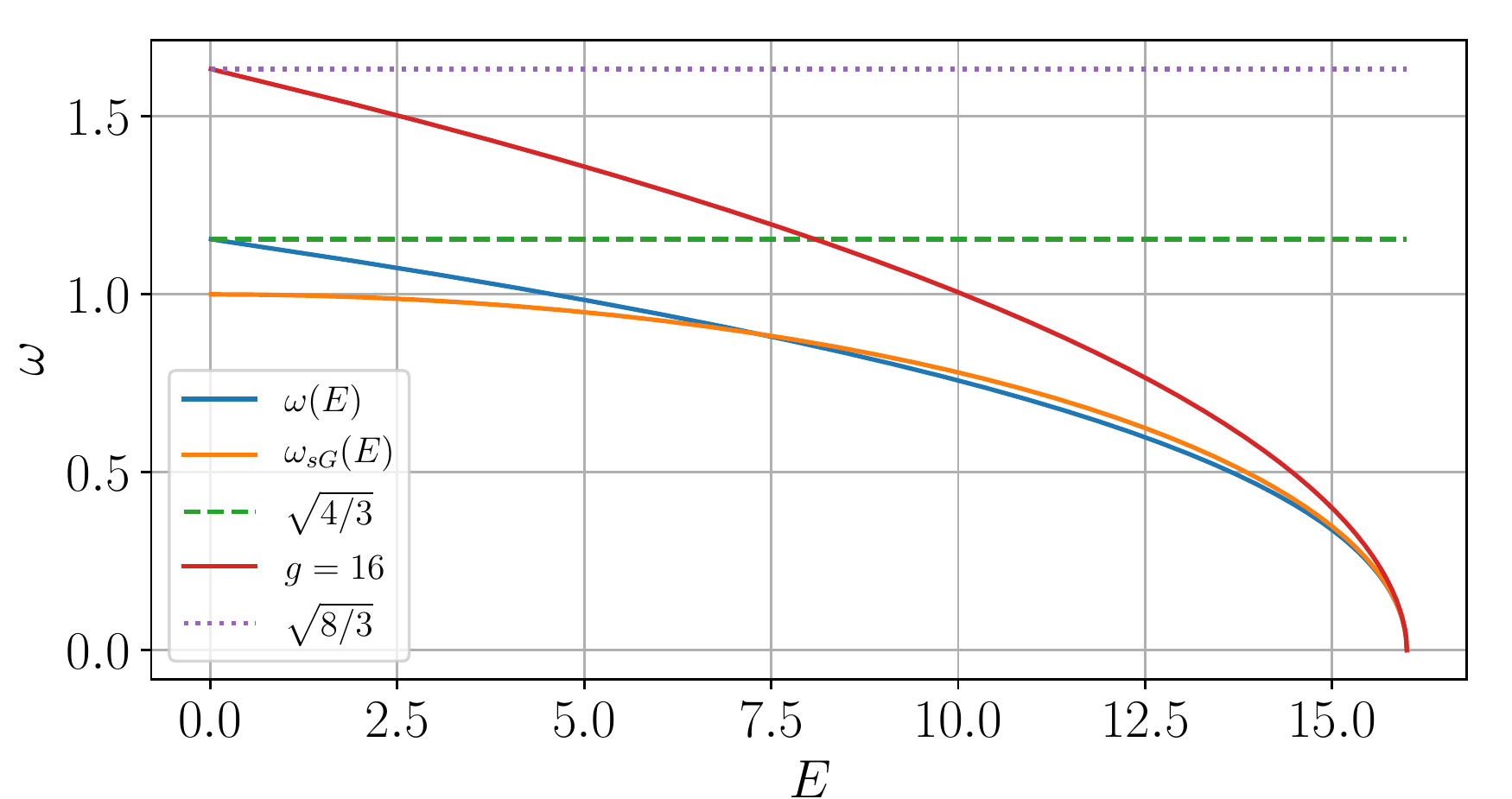}
\caption{Frequency vs. energy for the sine-Gordon breather,
$\omega_{\rm sG}(E)$, and for the periodic solutions of the moduli 
space dynamics, $\omega(E)$. The upper curve (red) shows the moduli
space result if the metric coefficient is fixed at its asymptotic
value $g=16$.}
\label{energy-freq-sg}
\end{figure}
and the frequency $\omega$ is $2\pi$ divided by this.
In Fig. \ref{energy-freq} we plot $E/\sqrt{1-\omega^2}$ for the moduli space
dynamics, and compare with the ratio $16$ for the exact sG
breather. The ratios agree as the energy $E$
approaches $16$ (i.e., as $a_{\rm max} \to \infty$). In
Fig. \ref{energy-freq-sg} we plot $\omega$ against 
$E$ for the exact breather and for the oscillating solutions in 
the moduli space over a larger range. Again the agreement is good, 
especially as $ E \to 16$. The agreement is substantially worse if
the variation of the metric $g$ with $a$ is ignored in the moduli 
space dynamics (i.e., if it is fixed to its asymptotic value $g=16$). This 
is also shown. 

Note that the exact breather's frequency approaches $1$ as its 
amplitude goes to zero, whereas in the moduli space dynamics 
the limiting frequency is $\sqrt{\frac{4}{3}}$. The difference 
is because the moduli space dynamics uses a field profile of 
constant width, whereas the exact breather gets arbitrarily broad. 
The broadening can be interpreted as the analytic
continuation of the Lorentz contraction of a colliding kink and
antikink, and is ignored in the moduli space approach.  

\subsection{Kink-antikink-kink dynamics in sine-Gordon theory}

The kink-antikink-kink analogue of the naive superposition of a 
kink and antikink in sine-Gordon theory is
\be
\phi(x;b) = 4\arctan(e^{x+b}) - 4\arctan(e^x) + 4\arctan(e^{x-b}) \,.
\label{sGKK*K}
\ee
To simplify, we have restricted attention to configurations
that are (anti)symmetric in $x$, in the sense that $\phi(-x) = 2\pi -
\phi(x)$. Using the tangent addition formula
\be
\tan(\alpha + \beta + \gamma) = 
\frac{\tan(\alpha) + \tan(\beta) + \tan(\gamma) 
- \tan(\alpha)\tan(\beta)\tan(\gamma)}
{1 - \tan(\alpha)\tan(\beta) - \tan(\alpha)\tan(\gamma) 
- \tan(\beta)\tan(\gamma)}
\ee
we can rexpress \eqref{sGKK*K} as
\be
\phi(x;b) = 4\arctan\left(
\frac{(2\cosh(b) - 1)e^x + e^{3x}}{1 +  (2\cosh(b) - 1)e^{2x}} \right)
\,.
\label{sGKK*Ksimp}
\ee
This configuration is symmetric in $b$, and $2\cosh(b) - 1$ is nowhere 
less than $1$. When $b=0$ it becomes a single kink.

\begin{figure}
\includegraphics[width=0.49\textwidth]{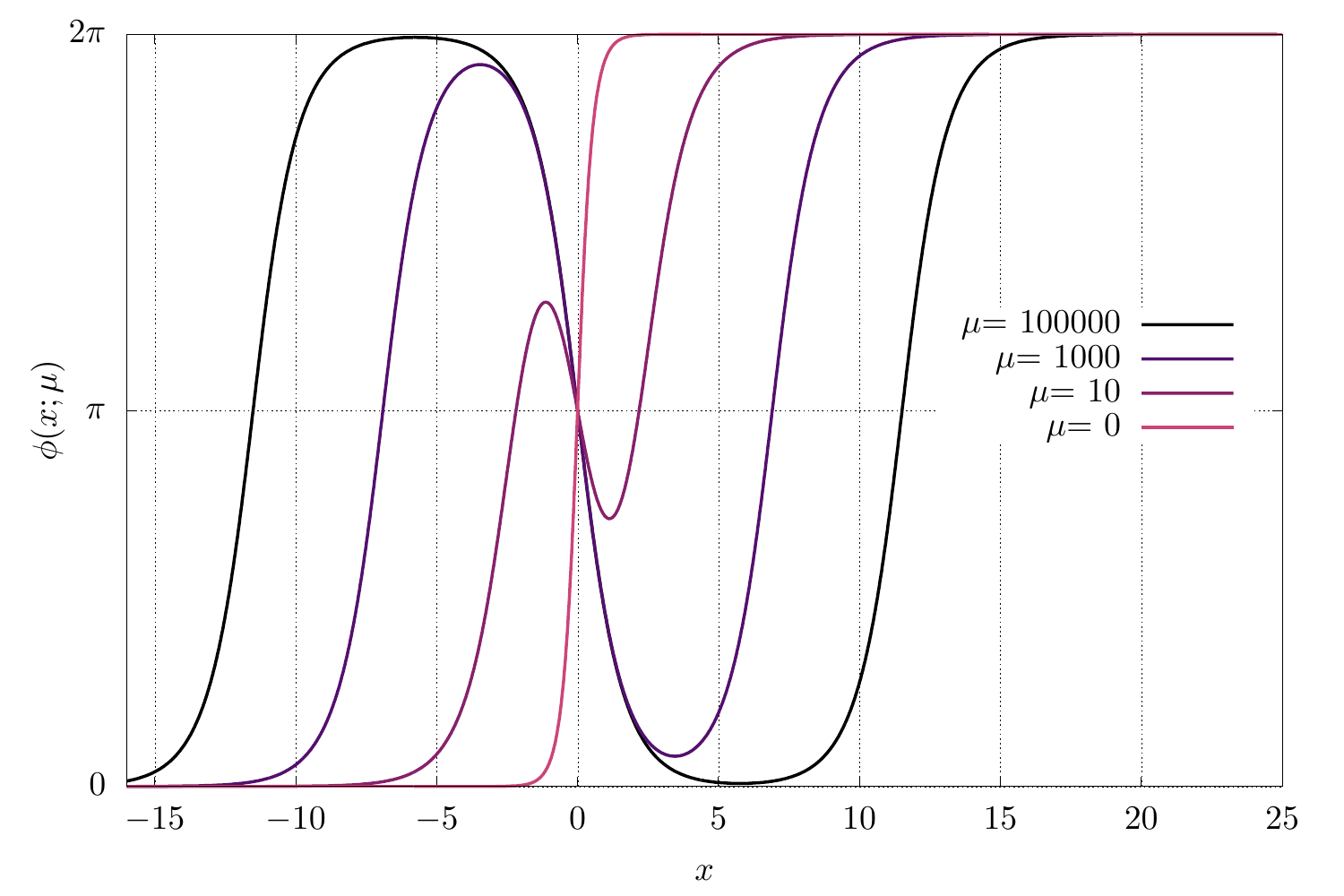}
\includegraphics[width=0.49\textwidth]{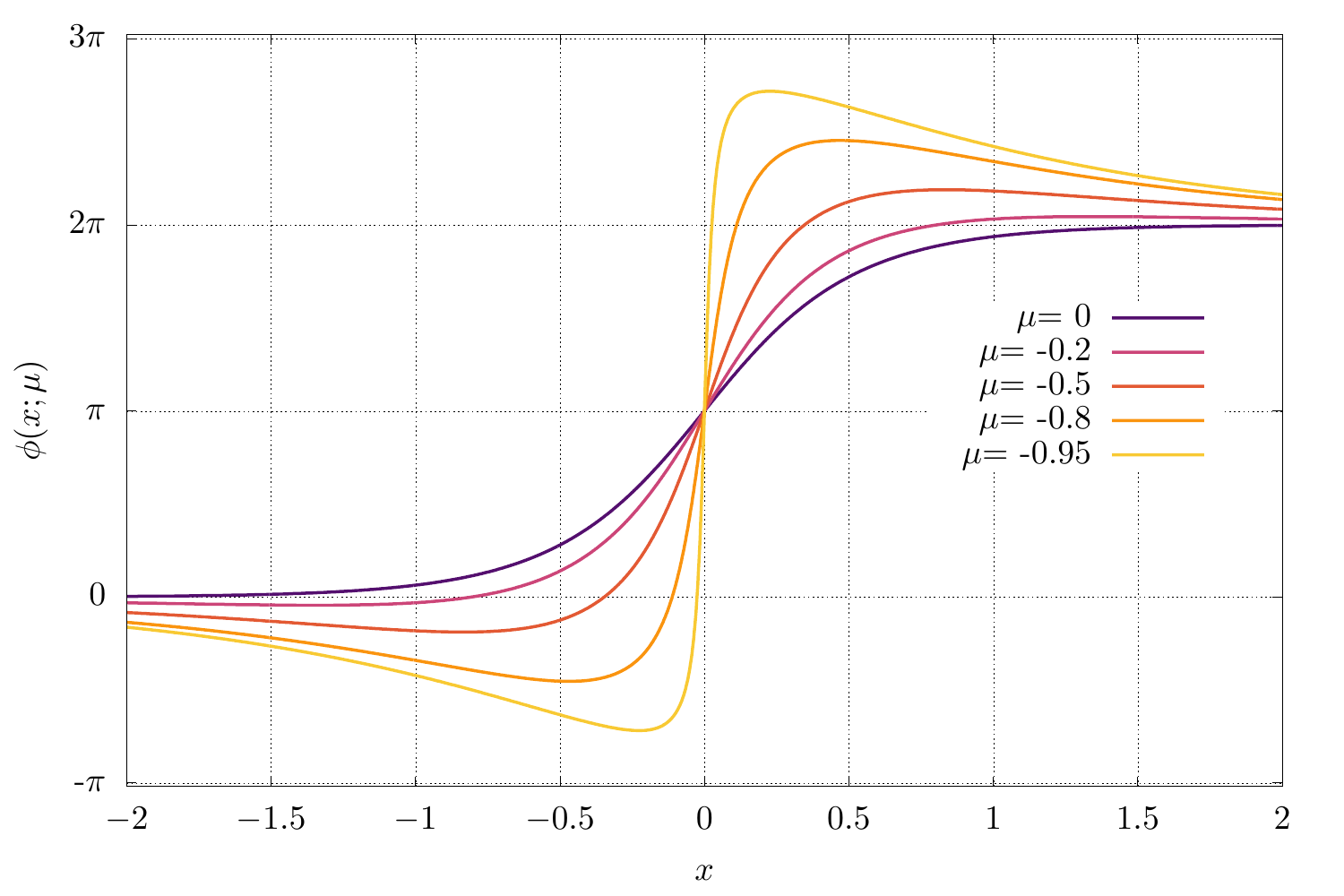}
\caption{ Examples of kink-antikink-kink configurations (\ref{sGKK*Kmu}).}
\label{sG-3-naive-mu}
\end{figure}

For these kink-antikink-kink configurations, the
derivative $\frac{\pr \phi}{\pr b}$ vanishes at $b=0$, so the metric
coefficient $g(b)$ on the moduli space vanishes there too. The 
moduli space is incomplete. This problem is resolved by extending 
to the set of configurations  
\be
\phi(x;\mu) = 4\arctan\left(
\frac{\mu e^x + e^{3x}}{1 + \mu e^{2x}} \right) \,,
\label{sGKK*Kmu}
\ee
where $\mu$ takes any real value greater than $-1$. For 
large positive $\mu$ we have a chain of kink, antikink and
kink, and for $\mu = 1$ a single kink. When $\mu = 0$ the
configuration is a compressed kink $\phi(x) = 4\arctan(e^{3x})$, and when 
$\mu$ is negative the kink is compressed further and
acquires bumps outside the usual range of field 
values $[0,2\pi]$. The bumps extend down to $-\pi$ and up to $3\pi$ 
as $\mu \to -1$, and the configuration is asymptotically
a half-antikink on the left (interpolating between 0
and $-\pi$), a compressed double-kink in the middle (interpolating 
between $-\pi$ and $3\pi$) and another half-antikink on the right 
(interpolating between $3\pi$ and $2\pi$). Examples of these 
configurations are plotted in Fig. \ref{sG-3-naive-mu}.

\begin{figure}
\includegraphics[width=0.49\textwidth]{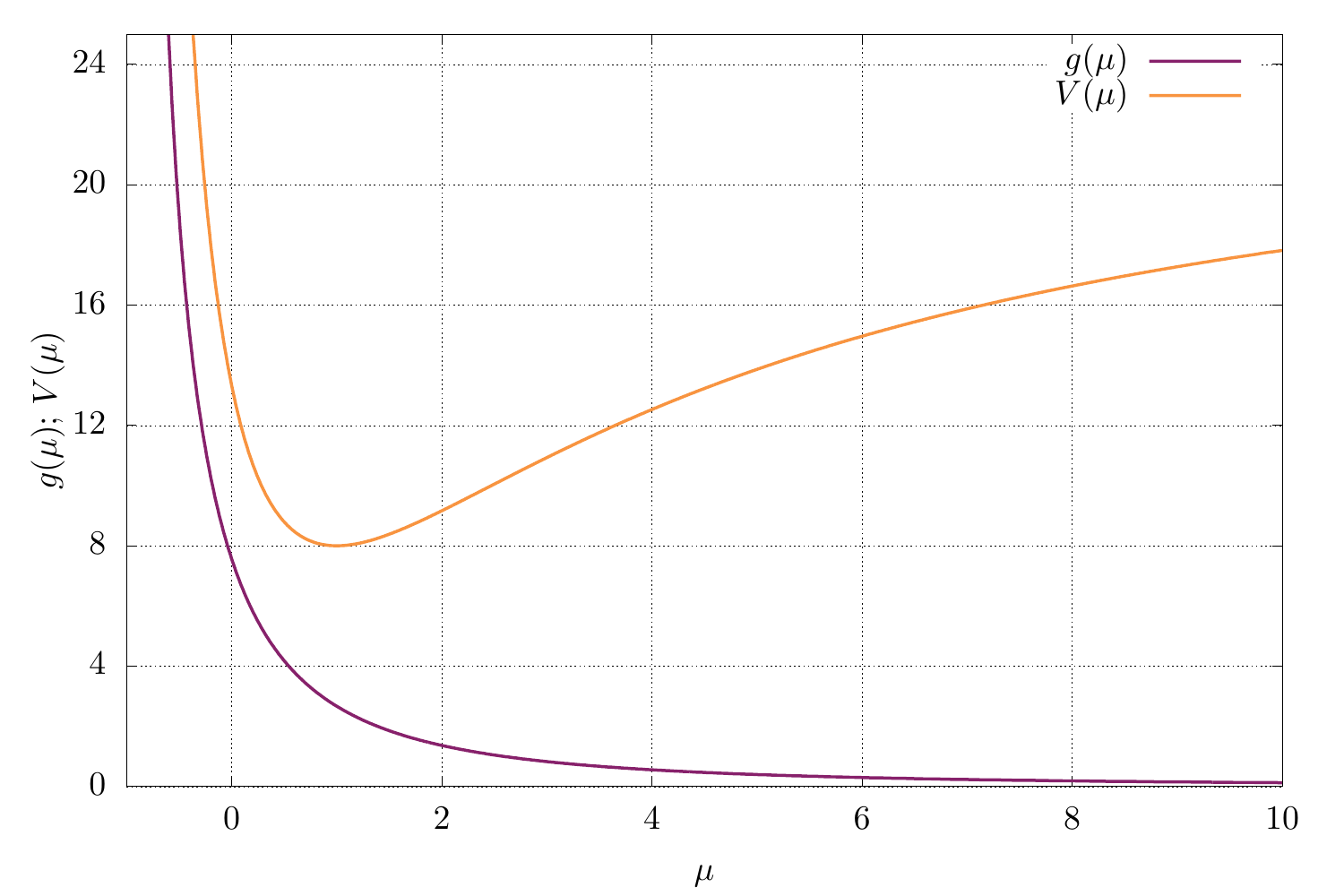}
\includegraphics[width=0.49\textwidth]{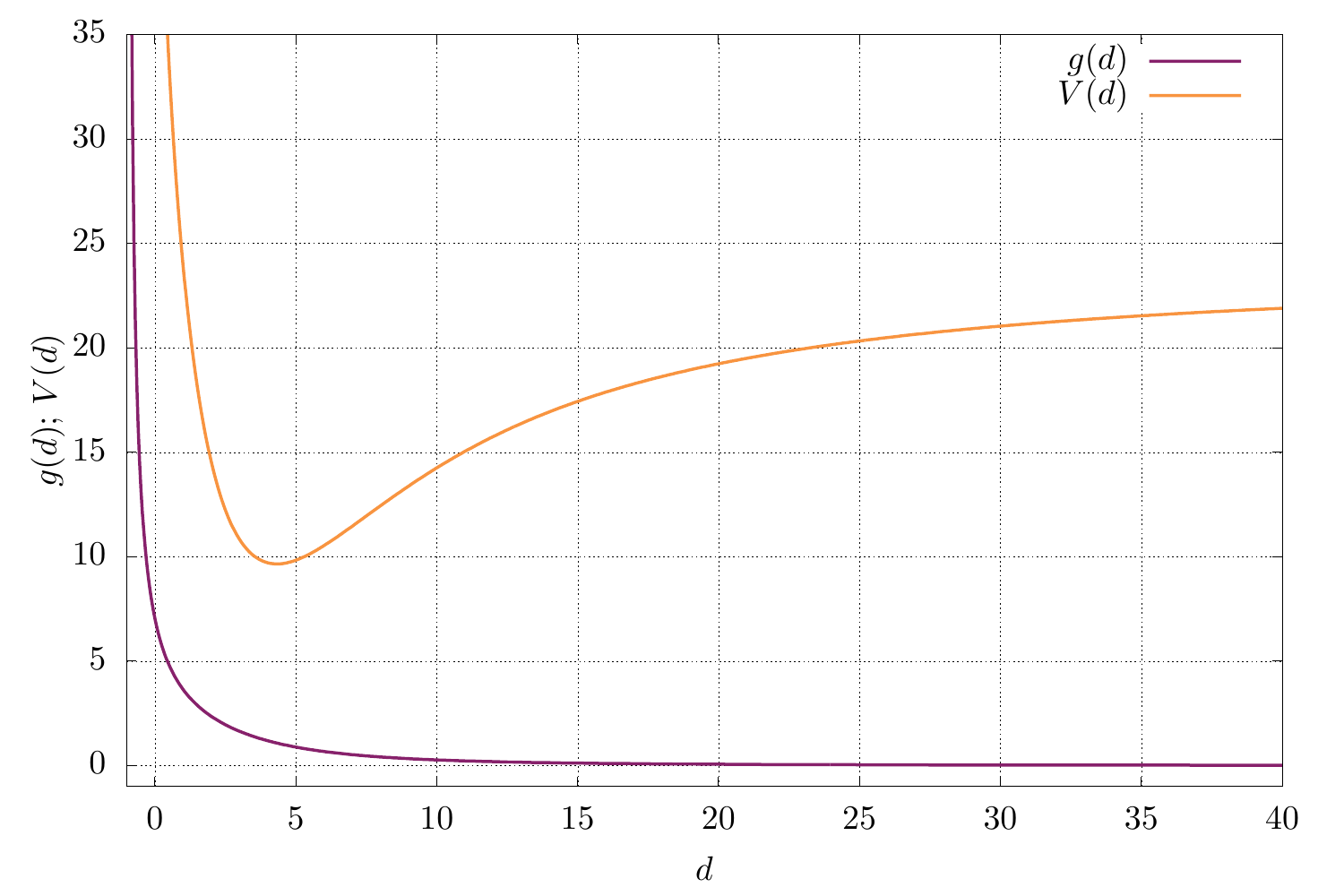}
\caption{The moduli space metric $g$ and potential $V$ for the
kink-antikink-kink configurations \eqref{sGKK*Kmu}
({\it left}) and \eqref{3-kink-d} ({\it right}).}
\label{sG-3-naive-metric}
\end{figure}

The metric on this moduli space, in terms of $\mu$, is 
\be
g(\mu)=\frac{16}{(\mu-1)(\mu+3)} \left[
1-\frac{4}{(\mu+1)\sqrt{(\mu-1)(\mu+3)}} \mbox{arccoth} \left(
\frac{1+\mu}{\sqrt{(\mu-1)(\mu+3)}} \right)\right] \,.
\ee
It is smooth and positive definite, but becomes 
singular as $\mu \to -1$. The metric and potential are shown 
in Fig. \ref{sG-3-naive-metric} (left panel). The 
dynamics on the moduli space gives a reasonably good description 
of symmetric kink-antikink-kink motion in sG theory.

However the exact sG kink-antikink-kink solutions involve
somewhat different field configurations. There is an analogue of a
breather solution in this sector \cite{CQS},
\be
\phi(x,t) = 4\arctan\left(\frac
{(1+\beta)e^x + (1-\beta)e^{x+2\beta x} - 2\beta e^{\beta x}
\cos(\alpha t)}
{(1-\beta) + (1+\beta)e^{2\beta x} - 2\beta e^{x+\beta x}
\cos(\alpha t)}\right) \,,
\label{Cuebreather}
\ee
where the parameters $\alpha$ and $\beta$ must satisfy $\alpha^2 +
\beta^2 = 1$. This solution has frequency $\alpha$ and its amplitude
depends on $\beta$. When $\alpha \to 1$ and $\beta \to 0$
it reduces to a static single kink, and for small $\beta$
it is known as a wobbling kink, or wobbler.
\begin{figure}
\includegraphics[width=0.49\textwidth]{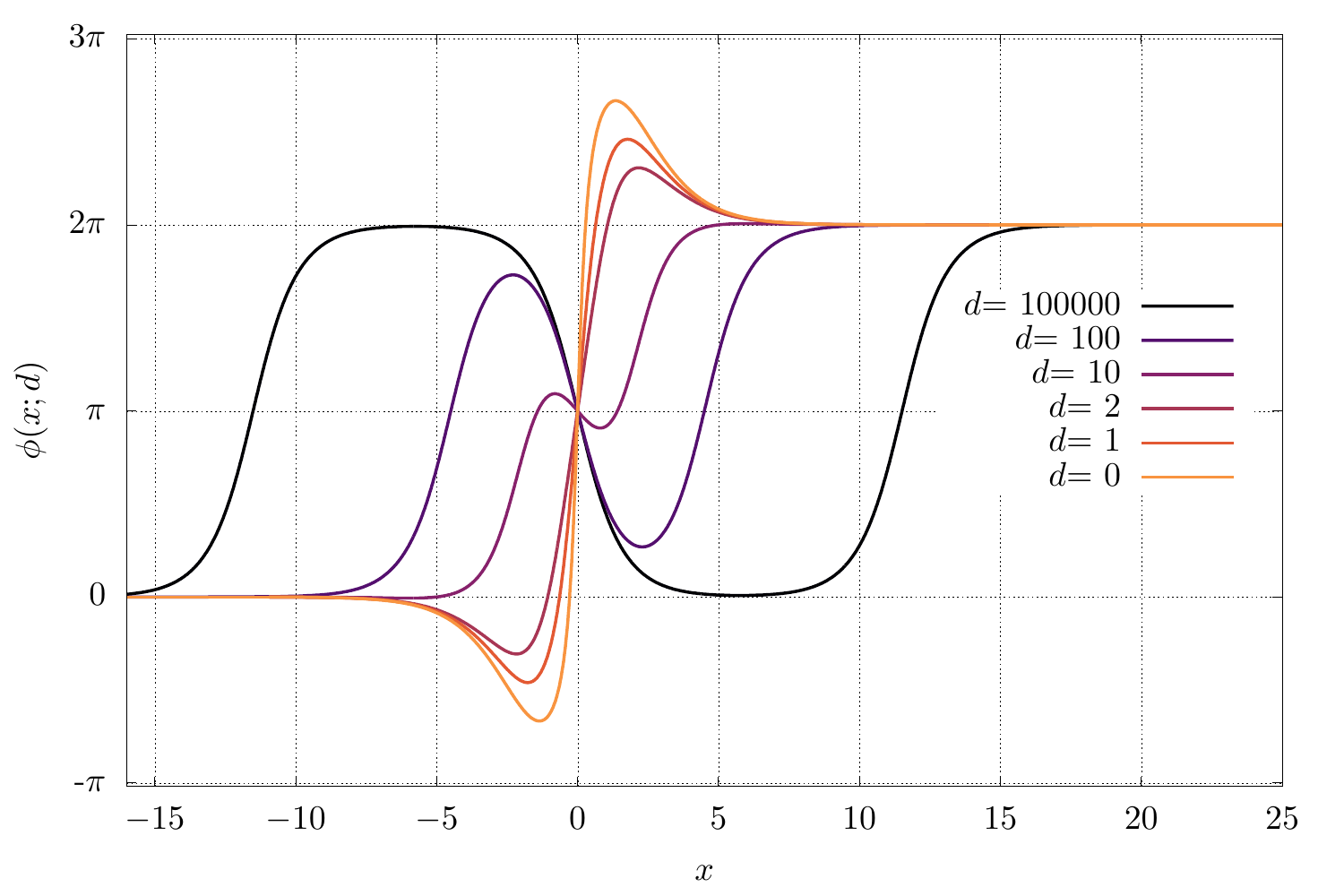}
\includegraphics[width=0.49\textwidth]{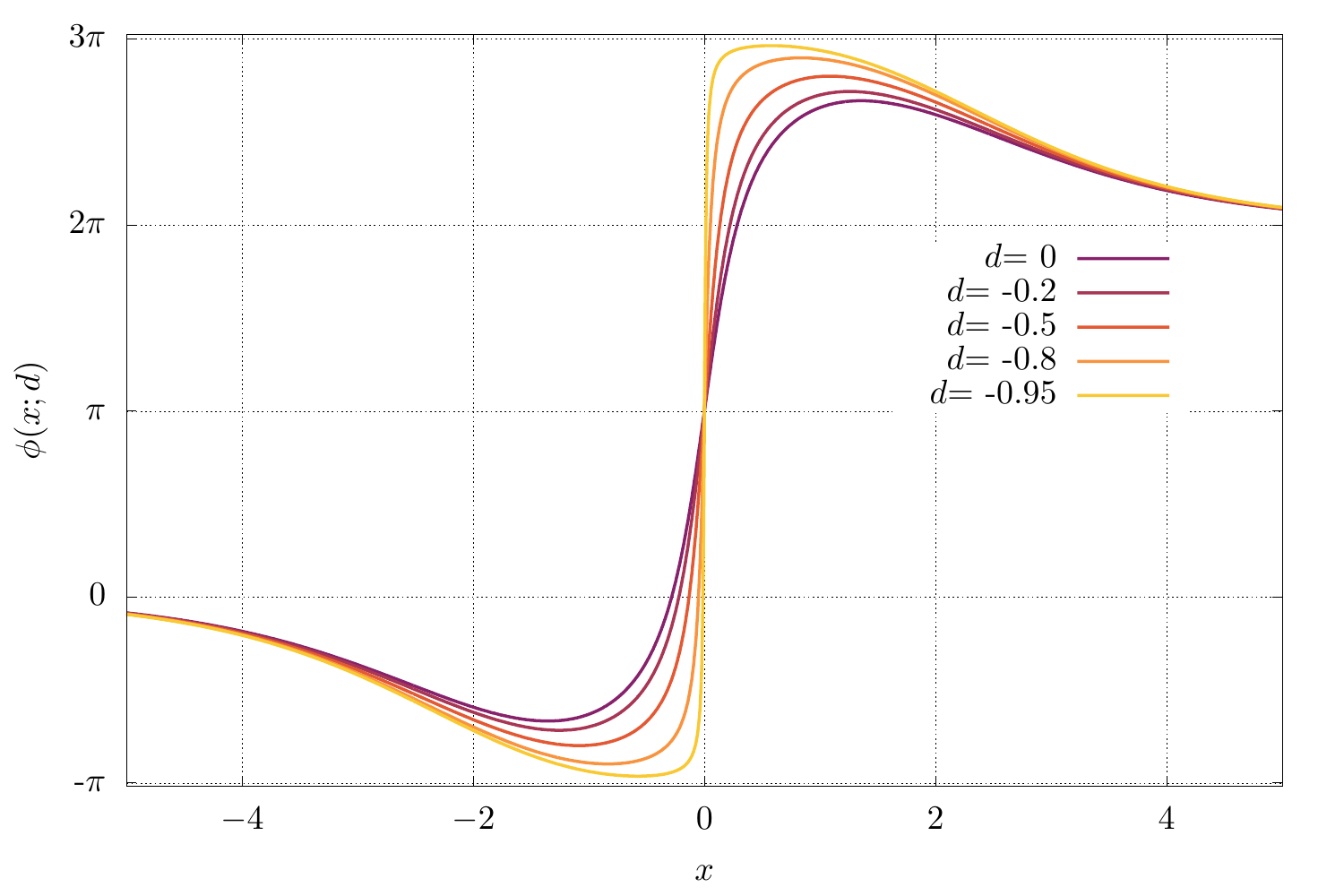}
\caption{Examples of kink-antikink-kink configurations (\ref{3-kink-d}).}
\label{sG-3-naive-d}
\end{figure}
We are interested in the opposite limit, where $\alpha \to 0$ 
and $\beta \to 1$. Taking this limit carefully, one finds the
analogue of the critical breather with infinite period, the solution
\be
\phi(x,t) = 4\arctan\left(\frac{(1+2t^2+2x)e^x + e^{3x}}
{1 + (1+2t^2-2x)e^{2x}}\right) \,.
\label{sGcritwob}
\ee
This has energy $E=24$, and approaches a configuration with
an infinitely separated kink, antikink and kink at rest as $|t| \to \infty$.

Notice that because of the terms $2x$ in formula \eqref{sGcritwob}, 
the field differs at all times from the configurations \eqref{sGKK*Kmu}. 
We therefore define a variant moduli space with configurations
\be
\phi(x;d) = 4\arctan\left(\frac{(d+2x)e^x + e^{3x}}
{1 + (d-2x)e^{2x}}\right) \,. \label{3-kink-d}
\ee
The solution \eqref{sGcritwob} moves precisely through this moduli space, but
only along the half-line $d \ge 1$. For the same reason as before, we
need $d$ to extend through zero to negative real values to 
have a metrically complete moduli space, and to accommodate dynamics with
energy greater than 24. The allowed range is again $d>-1$, because 
the configuration approaches an infinitely steep double-kink
sandwiched between two half-antikinks as $d \to -1$. Bumps 
occur here for positive and negative values of $d$, see
Fig. \ref{sG-3-naive-d}.

The metric $g(d)$ and potential $V(d)$ on the variant moduli space can be 
calculated in the usual way and are plotted in 
Fig. \ref{sG-3-naive-metric} (right panel). Because
the exact solution \eqref{sGcritwob} moves precisely 
through this moduli space, $g$ and $V$ are related. Combining the 
energy equation for the moduli space motion at $E=24$,
\be
\half g(d) \, {\dot d}^2 + V(d) = 24 \,,
\ee
with the time-dependence $d(t) = 1+2t^2$, implying that 
${\dot d}^2 = 16t^2 = 8(d-1)$, we obtain the relation
\be
4 g(d) (d-1) + V(d) = 24 \,.
\ee
\begin{figure}
\center \includegraphics[width=0.72\textwidth]{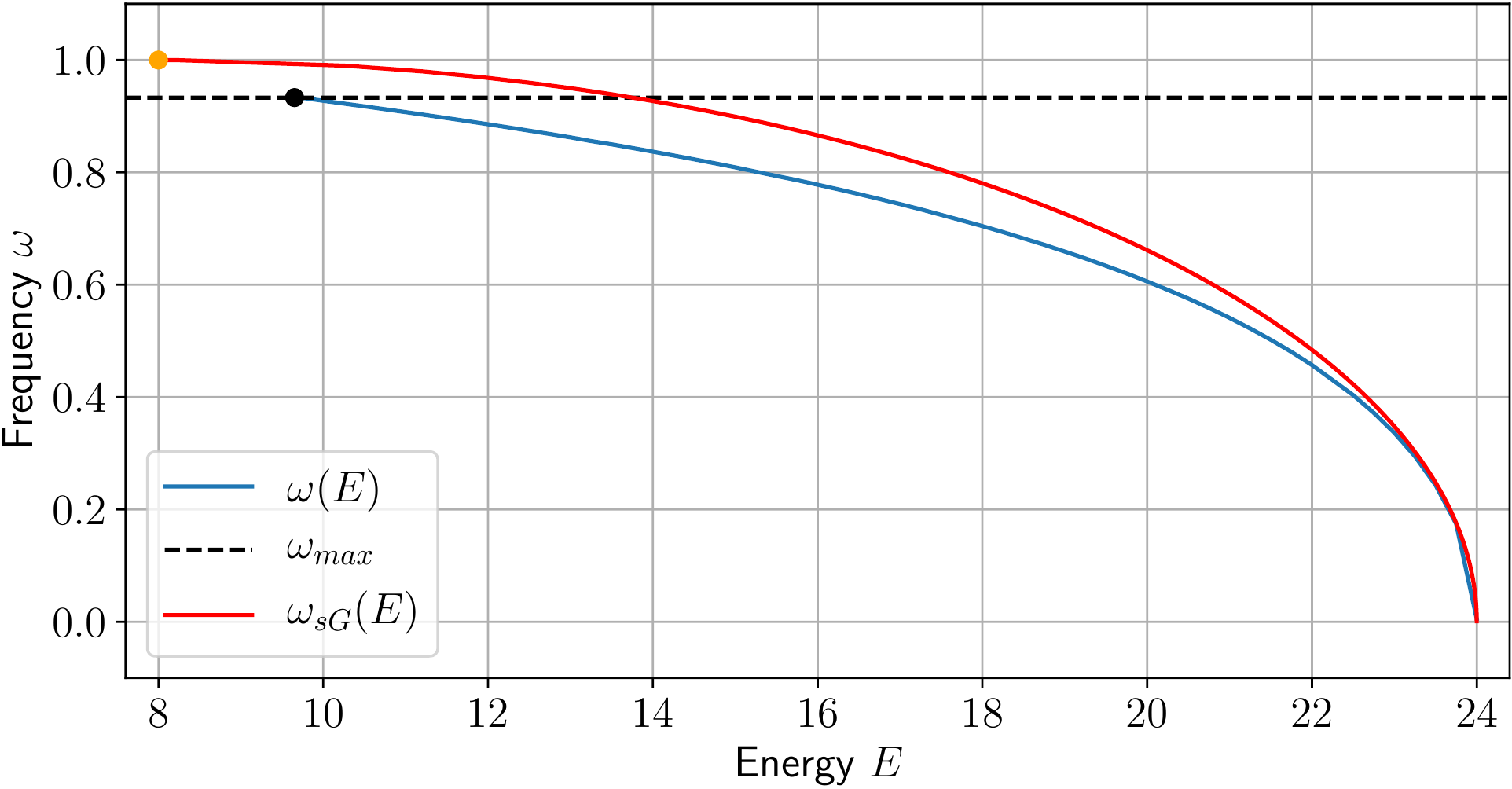}
\caption{Frequency vs. energy for the sine-Gordon wobbler, $\omega_{\rm sG}(E)$, 
and for periodic motion on the variant moduli space, $\omega(E)$.}
\label{energy-freq-3K}
\end{figure}

Motion on the variant moduli space with energy $E$ less than 24 gives
a good approximation to the wobbler. In Fig. \ref{energy-freq-3K} 
we plot frequency vs. energy, $\omega_{\rm sG}(E)$, for 
the exact wobbler and for its moduli space approximation 
$\omega(E)$. The curves agree quite well both for wobblers of
small amplitude, and for large wobblers, where the constituent 
solitons move further apart and the oscillating field stays close to
the exact solution \eqref{sGcritwob}.
Also, having extended the moduli space to negative $d$, we can model
a symmetrical collision of two incoming kinks with an antikink, with
energy greater than 24. The exact solutions describing collisions 
can be derived from eq.\eqref{Cuebreather} by allowing $\beta$ to 
be greater than $1$ and $\alpha$ imaginary, and they are well 
approximated by the moduli space dynamics.
 
This completes our discussion of moduli space dynamics for kinks and antikinks 
in sine-Gordon theory. Having a nontrivial metric and potential on 
the moduli space improves on the results obtained using Newtonian dynamics 
and the asymptotic kink-antikink force. We now return to $\phi^4$ theory.
 
\section{Kink-Antikink Moduli Spaces in $\phi^4$ Theory}

In this section we discuss more than one moduli space for kink-antikink
configurations in $\phi^4$ theory. We clarify their global geometry,
and how to choose coordinates. Kink-antikink dynamics has been much 
studied, see e.g. \cite{Sug, Mosh, CSW, KG, no-force, PLTC}, and only
one of our moduli spaces is novel.

\subsection{Naive superposition of kink and antikink}

The naive superposition of a kink and antikink is a field
configuration of the form
\be
\phi(x;a) = \tanh(x+a) - \tanh(x-a) - 1 \,.
\label{K*Knaive}
\ee
For large or modest positive values of $a$ this represents a kink
centred at $-a$ and an antikink at $a$. The field is symmetric in $x$,
and the centre of mass is at the origin. The field shift by
$-1$ is a little awkward, but essential to satisfy the boundary
condition $\phi \to -1$ as $x \to \pm\infty$. Note that $\chi(x) =
\tanh(x+a) - \tanh(x-a)$ is a linear field of the type we mentioned
earlier, approaching zero at spatial infinity.

The configurations (\ref{K*Knaive}), possibly modified to give the
kink and antikink some velocity, have often been used before, for
example as initial data in numerical simulations of 
kink-antikink collisions. By calculating the potential energy or 
the energy-momentum tensor for $a \gg 0$, one can find the 
force between a well-separated kink and antikink.  

\begin{figure}
\includegraphics[width=0.49\textwidth]{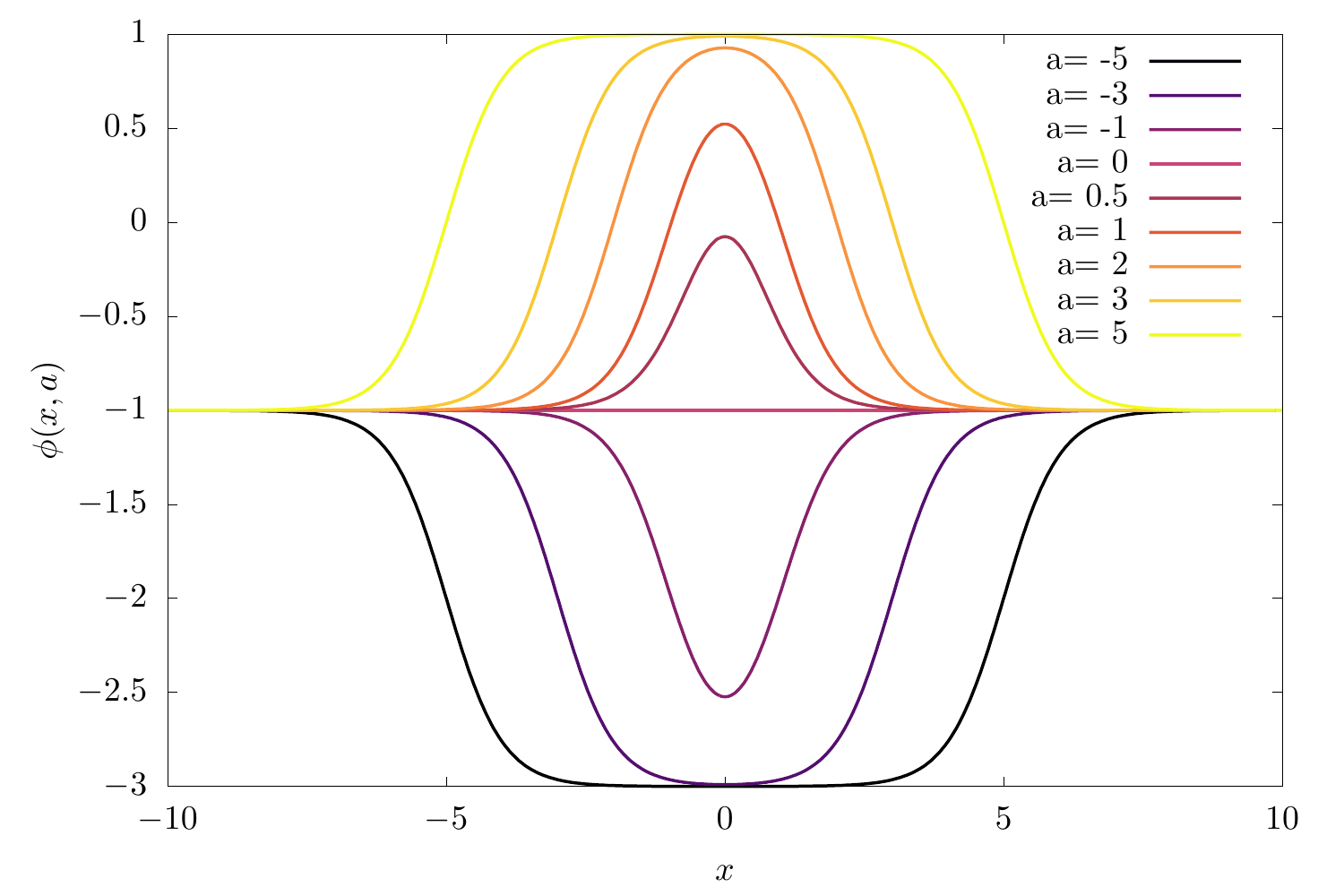}
\includegraphics[width=0.49\textwidth]{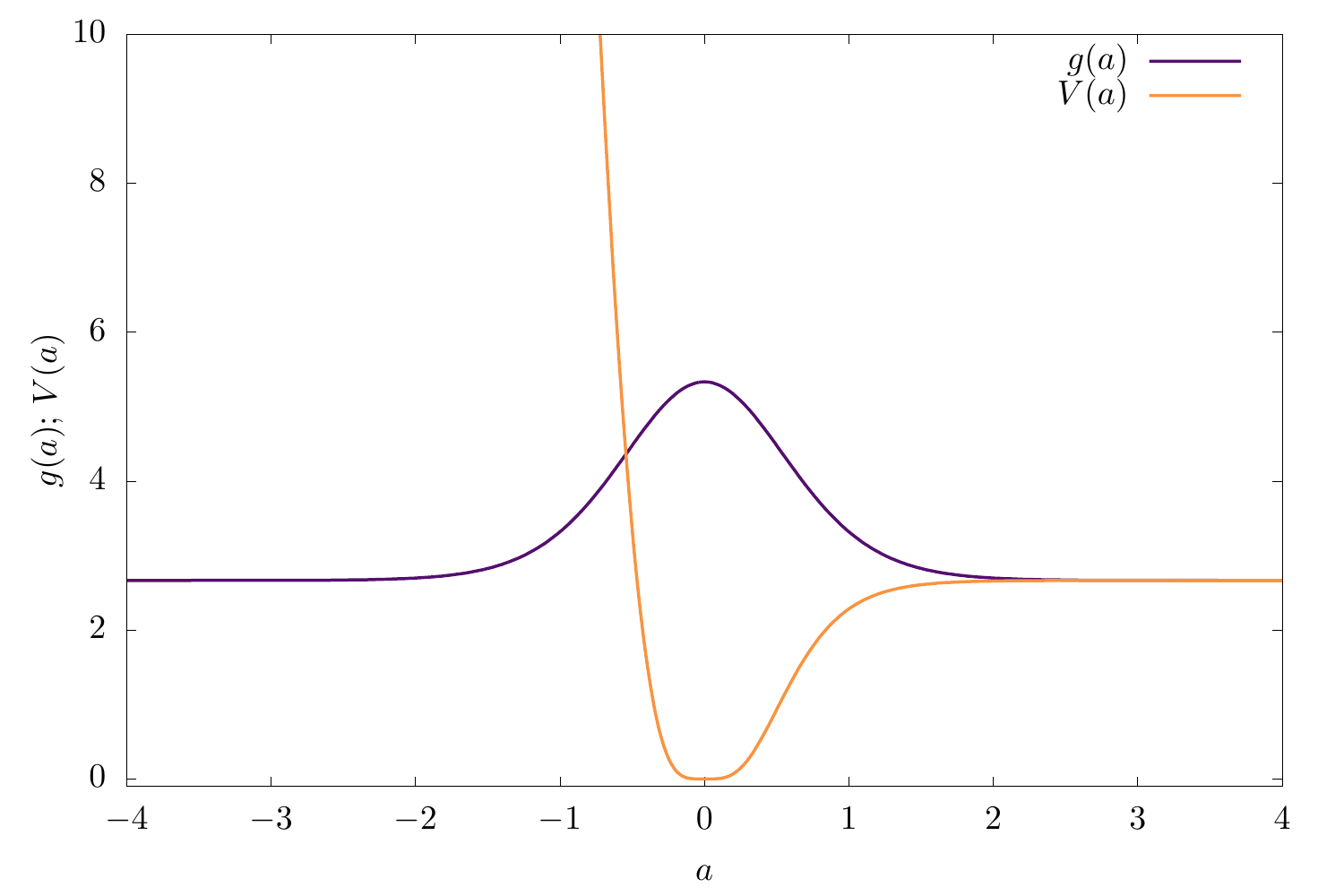}
\caption{{\it Left:} Naive superposition of a $\phi^4$ kink and antikink
for several values of the modulus $a$. {\it Right:} The
corresponding moduli space metric and potential.}
\label{naive-plot}
\end{figure}

The space of configurations $\phi(x;a)$ is a candidate 1-dimensional 
moduli space for a kink-antikink pair with their centre of mass fixed. 
The modulus $a$ runs along the real line $\R$. When $a$ is positive,
the kink is on the left and the antikink on the right. When $a = 0$, the
configuration is exactly the vacuum $\phi = -1$. As $a$ becomes
negative, the configuration $\phi(x;a)$ is less familiar, but note that
$\tanh(x+a) - \tanh(x-a)$ is antisymmetric in $a$ so as $a$ passes
through zero, the kink and antikink in some sense pass through each
other. For $a$ large and negative, the field interpolates from $-1$ to
$-3$ close to $x=a$ and back to $-1$ near $x=-a$. Configurations for
various values of $a$ are shown in Fig. \ref{naive-plot} (left panel).

The potential energy $V(a)$ for field configurations of the
form (\ref{K*Knaive}) is
\be
V(a)=\frac{8}{3} \frac{17+24a+9(-1+8a)e^{4a} -9e^{8a} +e^{12a}}
{(-1+e^{4a})^3} \,.
\ee
It sharply distinguishes positive and negative $a$. 
For $a \gg 0$ the energy $V$ is approximately $\frac{8}{3}$, twice
the kink mass, whereas for $a \ll 0$ it is dominated 
by the interval of length $2|a|$ where $\phi \approx -3$ and the 
energy density is approximately $32$; here, 
$V$ grows linearly with $|a|$. Dynamically there is no reason for
$\phi$ to prefer a value close to $-3$, so the field configurations that
occur in kink-antikink collisions, even at high speed, are probably
never close to configurations with large negative $a$ in this moduli space.
$V$ has its minimum at $a=0$, where the configuration is the vacuum.

To find the metric $g(a)$ on the moduli space, we need the derivative with
respect to $a$,
\be
\frac{\pr \phi}{\pr a}(x;a) = \frac{1}{\cosh^2(x+a)} +
\frac{1}{\cosh^2(x-a)} \,.
\label{Derivnaive}
\ee
This function is non-zero for all $a$, and symmetric in
$a$. Integrating its square, following eq.\eqref{metric-def}, we find
\be
g(a)=\frac{8}{\sinh^2(2a)} \left( -1 +\frac{2a}{\tanh{2a}} \right) 
+\frac{8}{3}\,.
\ee
The metric $g(a)$ and potential $V(a)$ are plotted in Fig. 
\ref{naive-plot} (right panel). Only the metric is symmetric.

It is simple to calculate $g(0)$. $\phi(x;a)$ has the 
Taylor expansion for $a \approx 0$,
\be
\phi(x;a) = -1 + \frac{2a}{\cosh^2(x)} + O(a^3) \,.
\ee
We call the function $\frac{1}{\cosh^2(x)}$ a bump, so the field
changes from having a positive bump to a negative bump (around $\phi
= -1$) as $a$ changes from positive to negative. At $a=0$, 
$\frac{\pr \phi}{\pr x} = \frac{2}{\cosh^2(x)}$, so
\be
g(0) = \int_{-\infty}^{\infty} \frac{4}{\cosh^4(x)} \, dx =
\frac{16}{3} \,,
\ee
eight times the gradient contribution to the kink
mass $M$, and therefore equal to $4M$.

In summary, the moduli space of the naive kink-antikink superposition is
metrically complete, and motion through this moduli space represents a kink
and antikink passing through each other smoothly from positive to
negative separation. The reduced Lagrangian on moduli space is
\be
L_{\rm red} = \half g(a) \, {\dot a}^2 - V(a) \,,
\ee
and the equation of motion has the first integral
\be
\half g(a) \dot{a}^2 + V(a) = E \,.
\ee
In a kink-antikink collision, $a$ decreases from a positive value and 
passes through zero, then stops at a maximal negative value determined by
energy conservation, and returns to a positive value. 
While $a$ is negative the field has a negative bump. 
The time taken for this bounce process can be estimated
by solving the equation of motion for $a$. However,
numerical exploration of the field equation for $\phi$ shows that this
moduli space is too simple to capture important features of 
kink-antikink dynamics. Most importantly, there is transfer of energy from the 
positional motions of the kink and antikink into shape mode oscillations. 
This can lead to multiple oscillations of $a$ before the kink and 
antikink emerge, or to kink-antikink annihilation. The
next moduli space therefore includes the shape mode.

\subsection{Including the shape mode}

A single kink, with profile $\phi(x) = \tanh(x)$, has a small-amplitude
shape mode of oscillation 
\be
\phi(x,t) = \tanh(x) + A(t)\frac{\sinh(x)}{\cosh^2(x)} \,,
\ee 
where $A(t)$ oscillates harmonically with
frequency ${\sqrt 3}$. This discrete frequency is below the frequencies
of continuum radiation modes which start immediately above 2.
Large amplitude oscillations of $A$ are not exact solutions of the field
equation. Energy is converted into radiation through a nonlinear
process, and $A$ slowly decays.

Configurations of a single kink, including this mode, are
\be
\phi(x;a,A) = \tanh(x-a) + A\frac{\sinh(x-a)}{\cosh^2(x-a)}\,.
\ee
The moduli space is 2-dimensional, with coordinates $a$ and $A$ both taking
values in all of $\R$. As
the derivatives of $\phi$ with respect to $a$ and $A$ are both non-zero
functions, and linearly independent, the metric is positive
definite. Its explicit form is
\be
g_{aa}=\frac{4}{3}  + \frac{\pi}{2} A + \frac{14}{15} A^2 \,, \quad
g_{aA}=0 \,, \quad g_{AA} = \frac{2}{3} \,.
\ee
Thus $a$ and $A$ are
globally good coordinates. The moduli space dynamics describes a kink
that oscillates indefinitely while also moving at a steady mean velocity. 

A set of field configurations for a kink at $-a$ and an antikink $a$ 
that includes this mode for each of them is
\be
\phi(x;a,A) =  \tanh(x+a) - \tanh(x-a)
+ A\left(\frac{\sinh(x+a)}{\cosh^2(x+a)} -
\frac{\sinh(x-a)}{\cosh^2(x-a)}\right) - 1 \,.
\ee
Again, the moduli space is 2-dimensional with coordinates $a$ and $A$.
The centre of mass is fixed at the origin, and the amplitudes of the
shape modes are correlated so that $\phi$ is symmetric in $x$. This
restricted set of fields is adequate for initial kink-antikink data having
the symmetry. The more general set of field configurations where
the shape mode amplitudes are independent was introduced by Sugiyama
\cite{Sug} and has been investigated by Takyi and Weigel \cite{TW} and
others. 

For a well-separated kink-antikink pair, it is clear that $a$ and $A$ are
good coordinates. $a$ can be positive or negative, and the shape mode
usefully allows a change of shape of both kink and antikink. However, at
$a=0$ there is a problem, because the shape modes cancel out and $A$
fails to be a good coordinate. The derivatives of $\phi$ with
respect to $a$ and $A$ need to be non-zero and linearly independent
functions for the metric on moduli space to be positive definite (in
these coordinates). This requirement is not satisfied at $a=0$ because
the derivative with respect to $A$ is zero. This problem was noted by
Takyi and Weigel as a null-vector problem.

The problem has a simple resolution by a change of coordinates. Replace
$A$ by $\frac{B}{f(a)}$ where $f(a)$ is any antisymmetric function of
$a$ that is linear for small $a$. The linear behaviour near
$a=0$ is important to remove the apparent metric singularity, and
allow for a smooth field evolution through $a=0$, but
otherwise the choice of $f$ is arbitrary. A change of $f$ 
corresponds to a change in the coordinate $B$, but the moduli space 
geometry is invariant. One could choose $f(a) = a$ but we
prefer $f(a) = \tanh(a)$, as $B$ is then the amplitude of the shape
mode, up to a sign, for large $|a|$. The moduli space of field 
configurations is now
\be
\phi(x;a,B) =  \tanh(x+a) - \tanh(x-a) + \frac{B}{\tanh(a)}
\left(\frac{\sinh(x+a)}{\cosh^2(x+a)} -
\frac{\sinh(x-a)}{\cosh^2(x-a)}\right) - 1 \,, 
\label{collective-aB}
\ee
and the function of which $B$ is the coefficient has a smooth, non-zero
limit as $a \to 0$. Using the Taylor expansion or equivalently
l'H\^opital's rule, one finds for small $a$ the leading terms
\be
\phi(x;a,B) \approx 2a \frac{1}{\cosh^2(x)} + 2B \left(
\frac{2}{\cosh^3(x)} - \frac{1}{\cosh(x)} \right) - 1 \,.
\ee
As $a$ and $B$ are coefficients of distinct non-zero functions, the
moduli space is non-singular in these coordinates and has a smooth
metric. More generally, $a$ and $B$ are globally good coordinates, 
taking values in all of $\R$.
\begin{figure}
\includegraphics[width=0.49\textwidth]{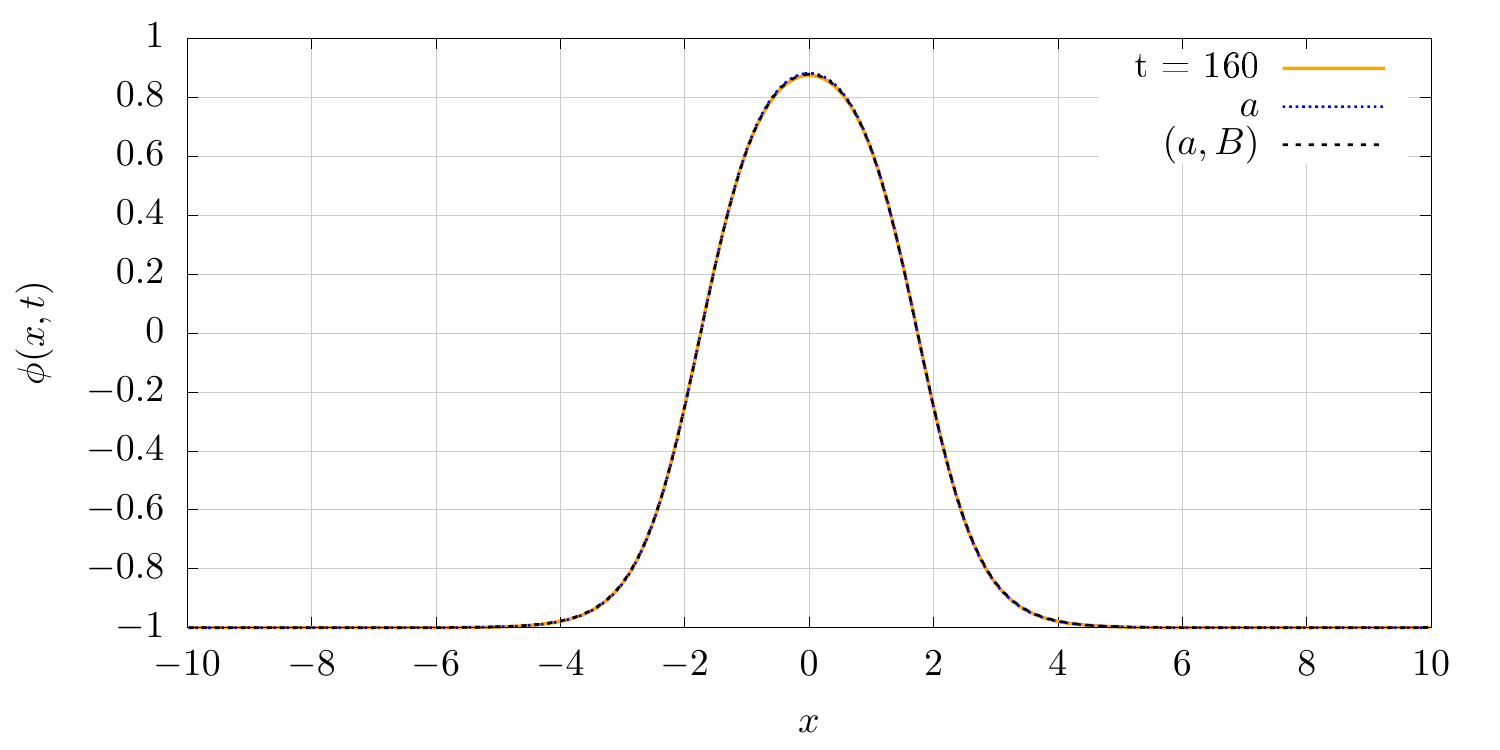} \includegraphics[width=0.49\textwidth]{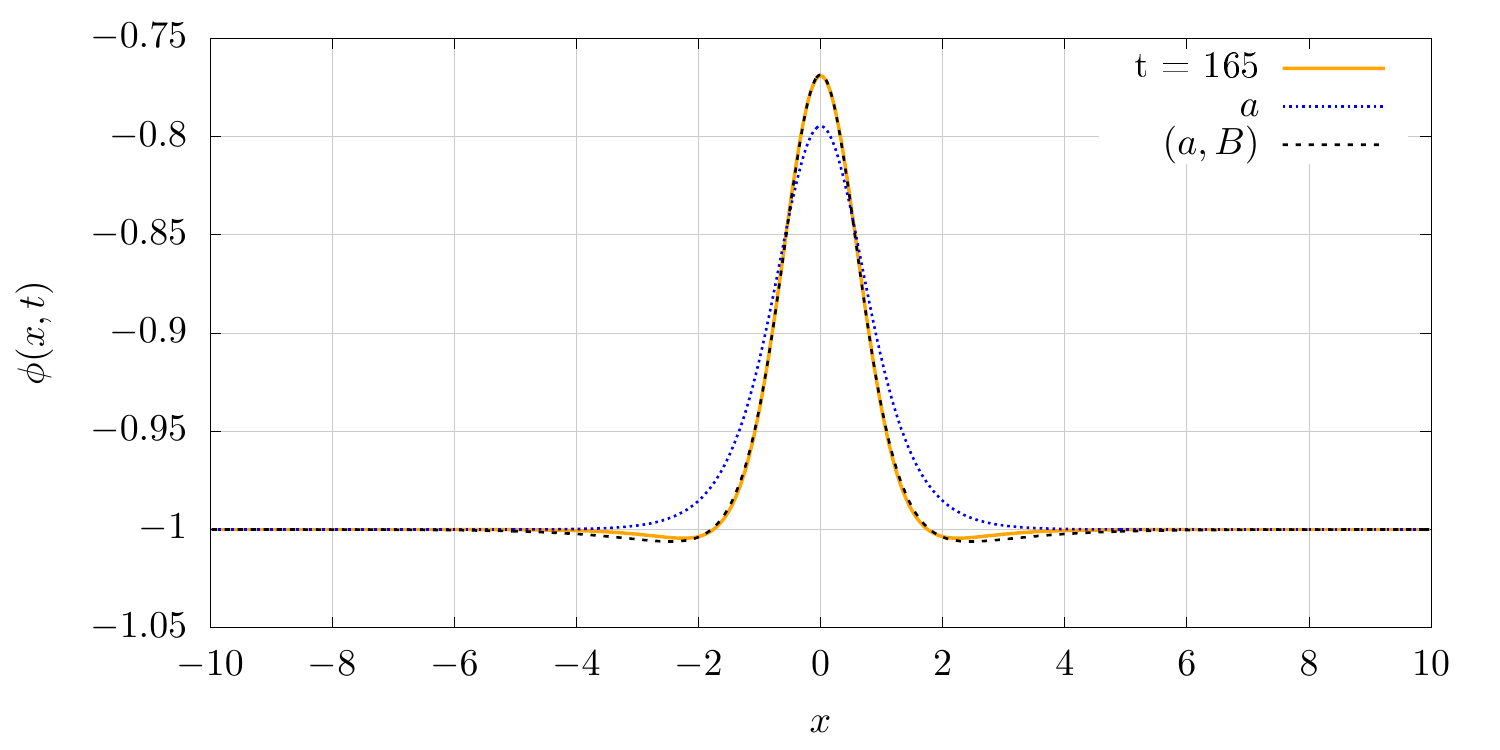}
\includegraphics[width=0.49\textwidth]{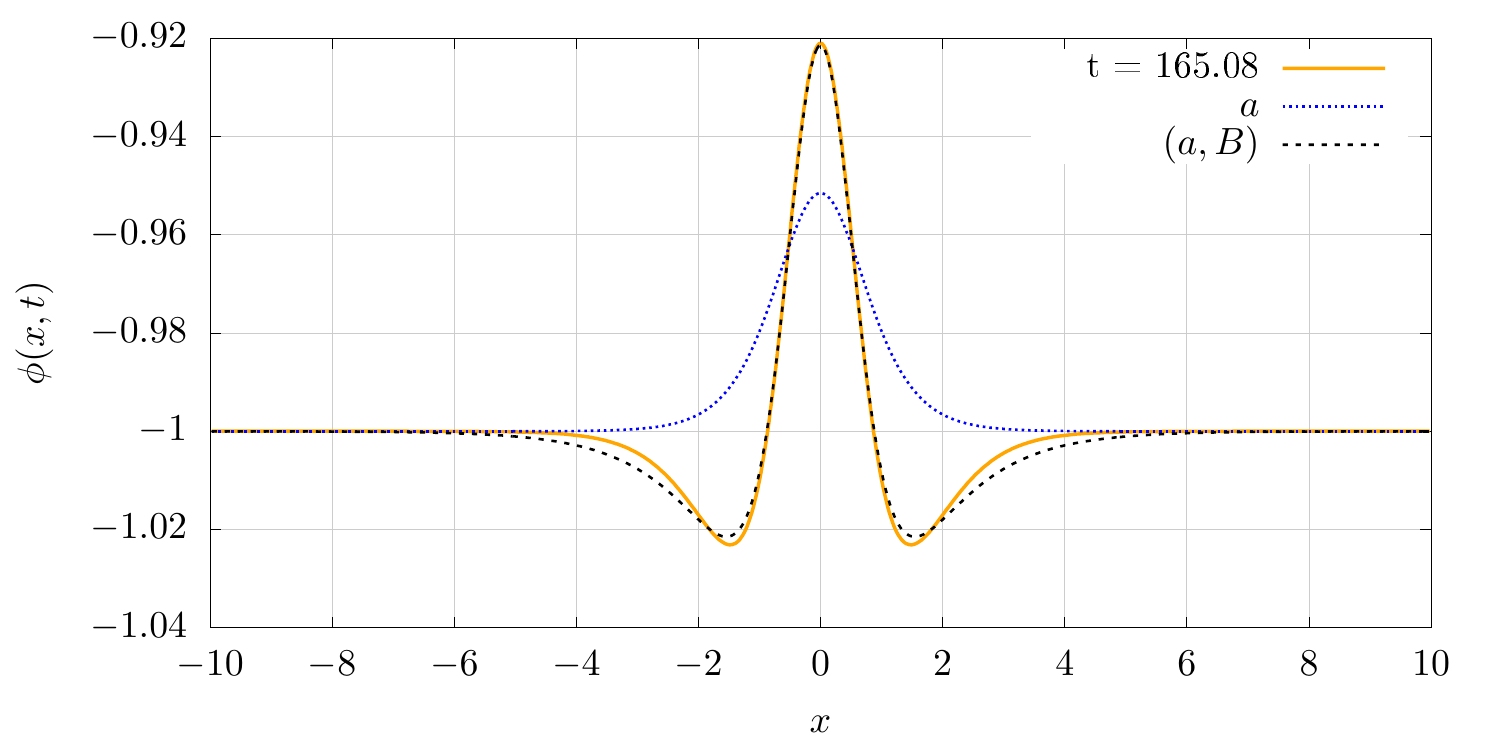} \includegraphics[width=0.49\textwidth]{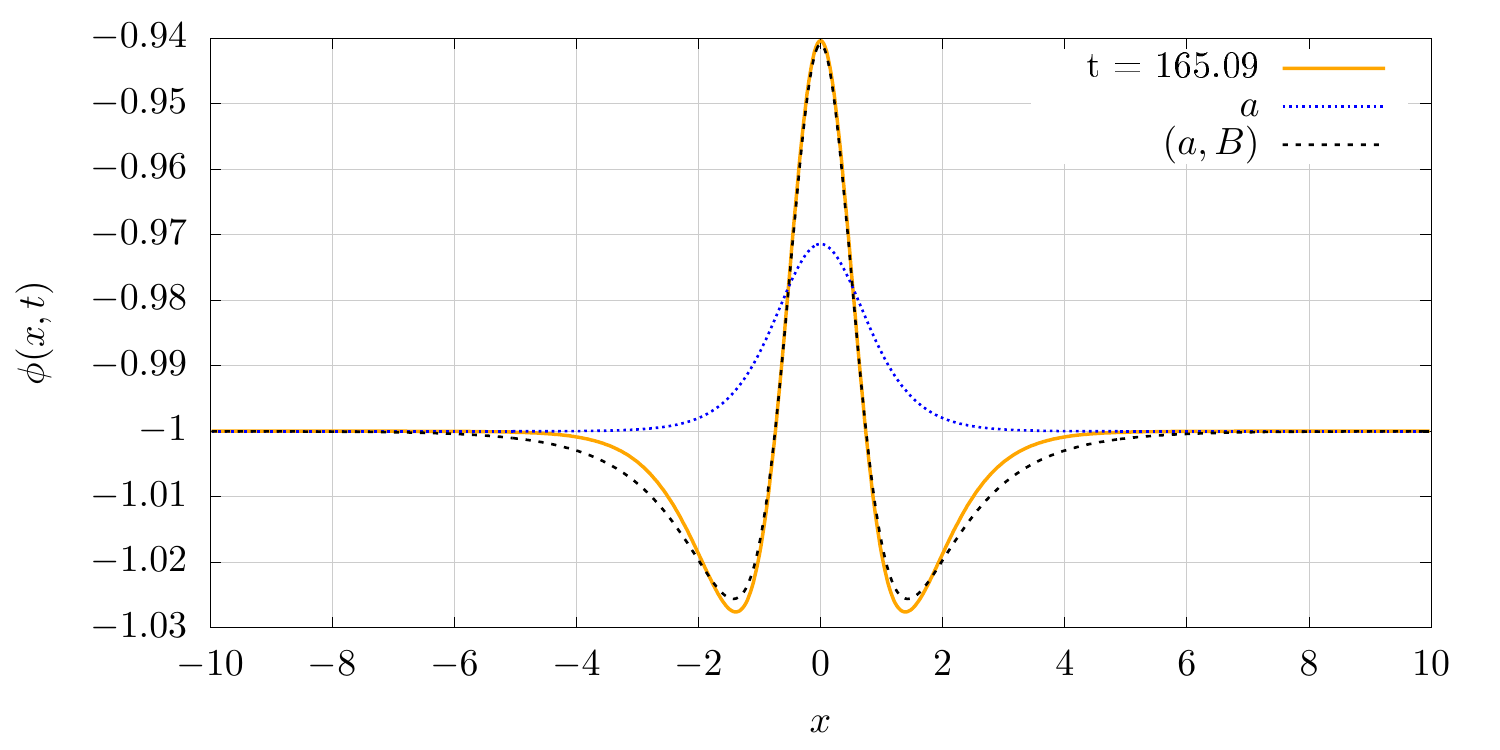}
\includegraphics[width=0.49\textwidth]{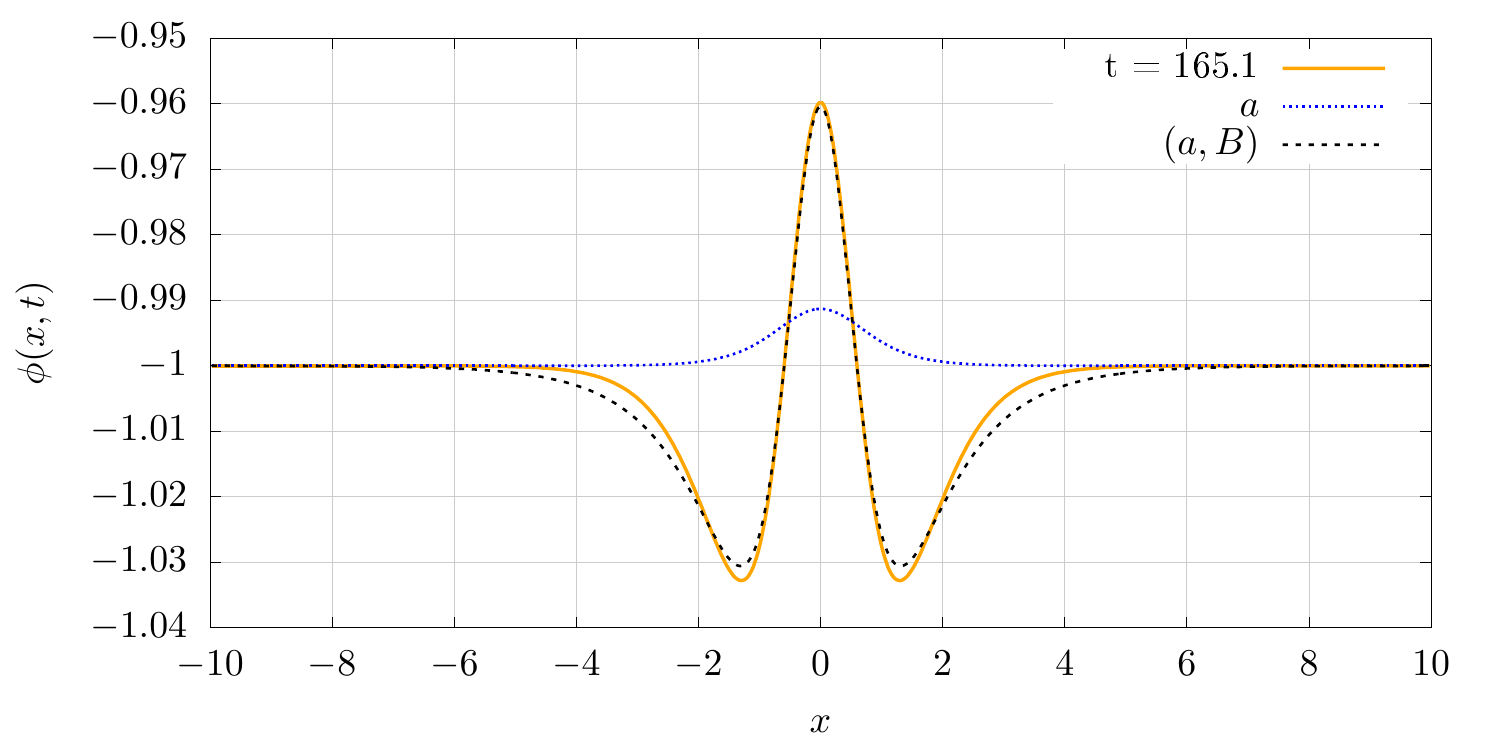} \includegraphics[width=0.49\textwidth]{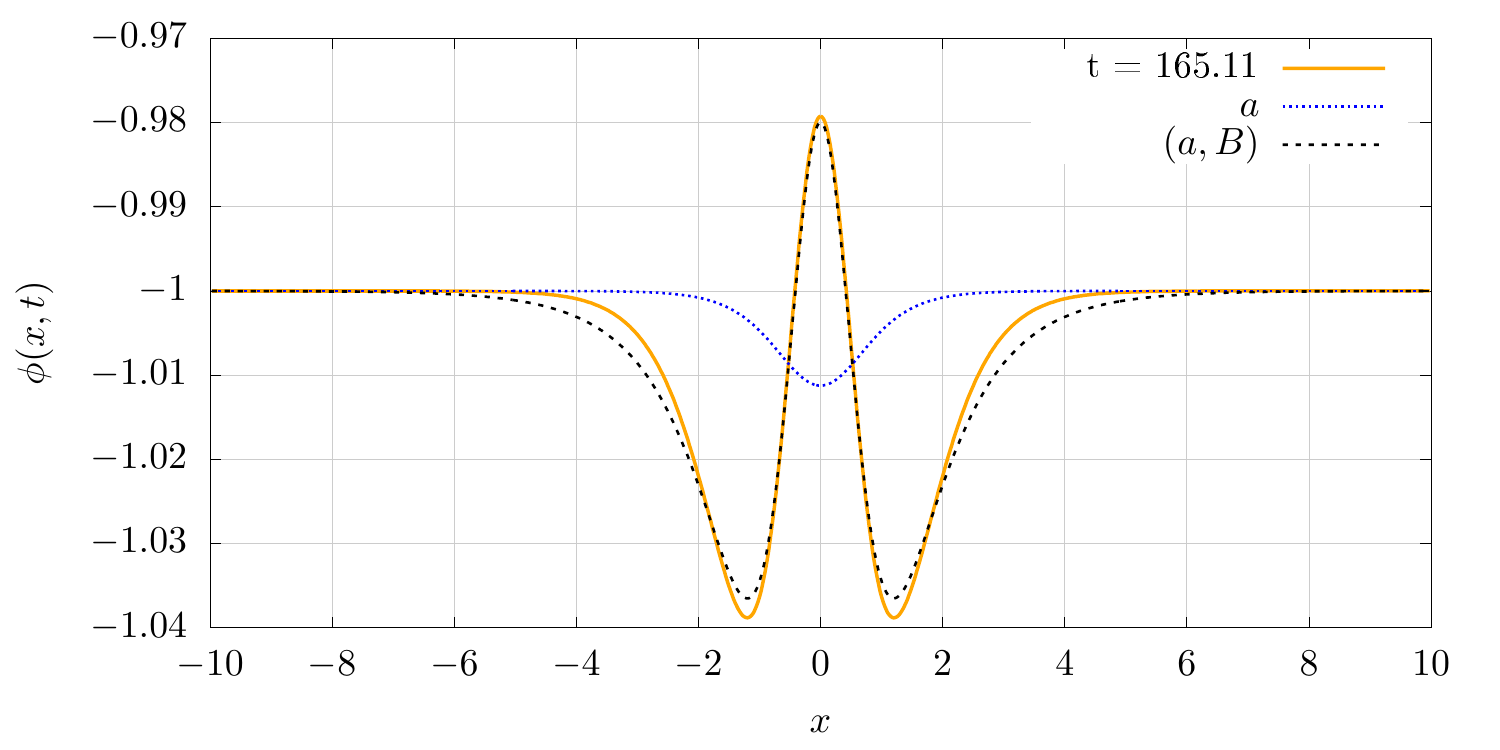}
\includegraphics[width=0.49\textwidth]{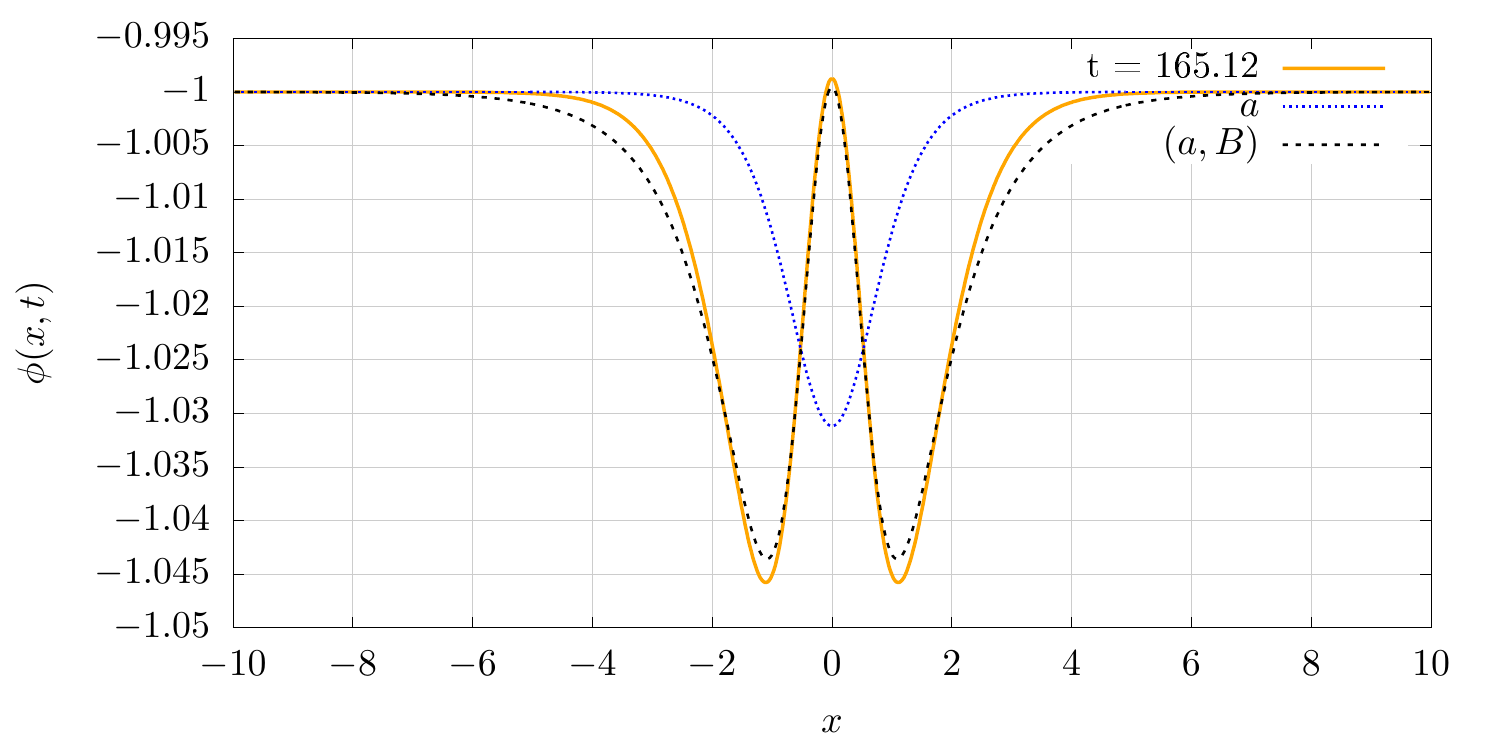} \includegraphics[width=0.49\textwidth]{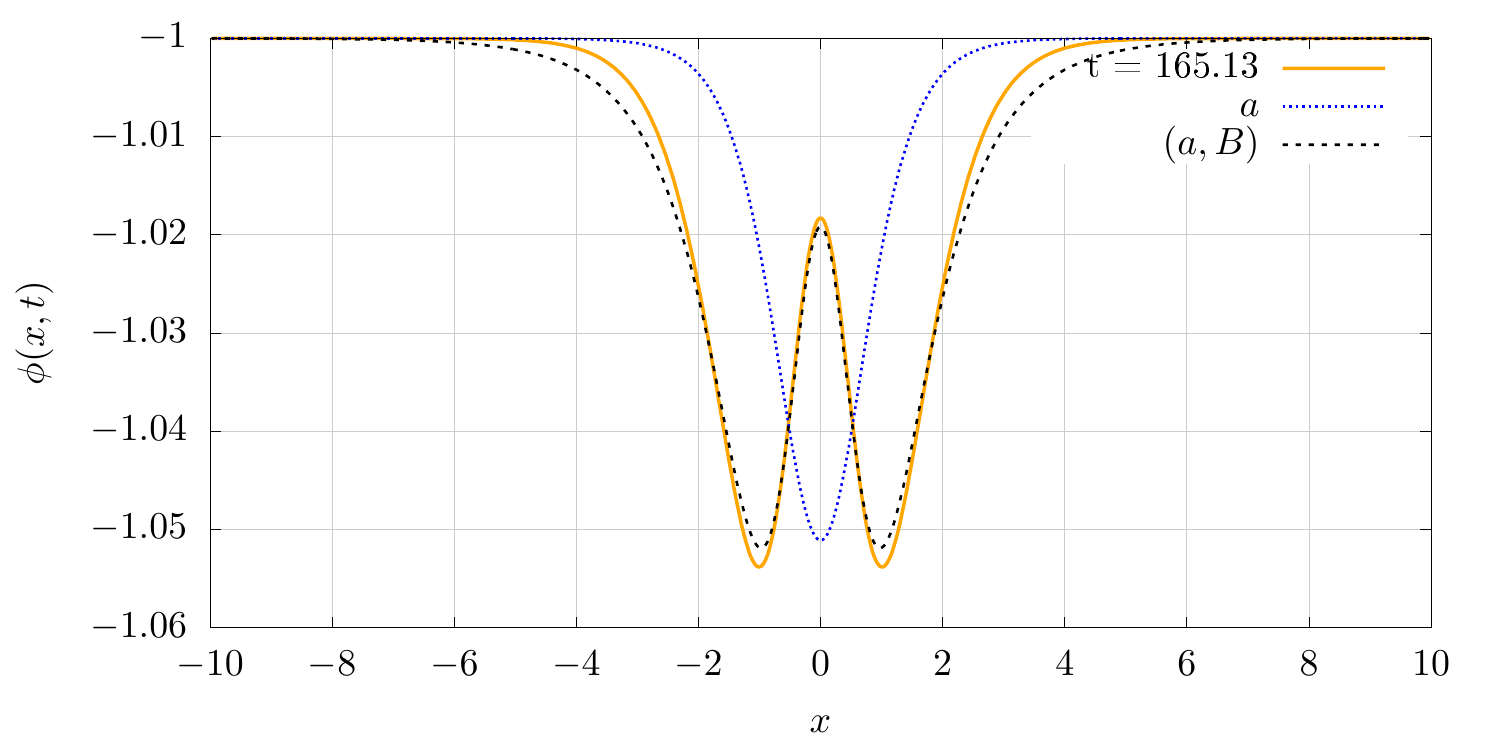}
\includegraphics[width=0.49\textwidth]{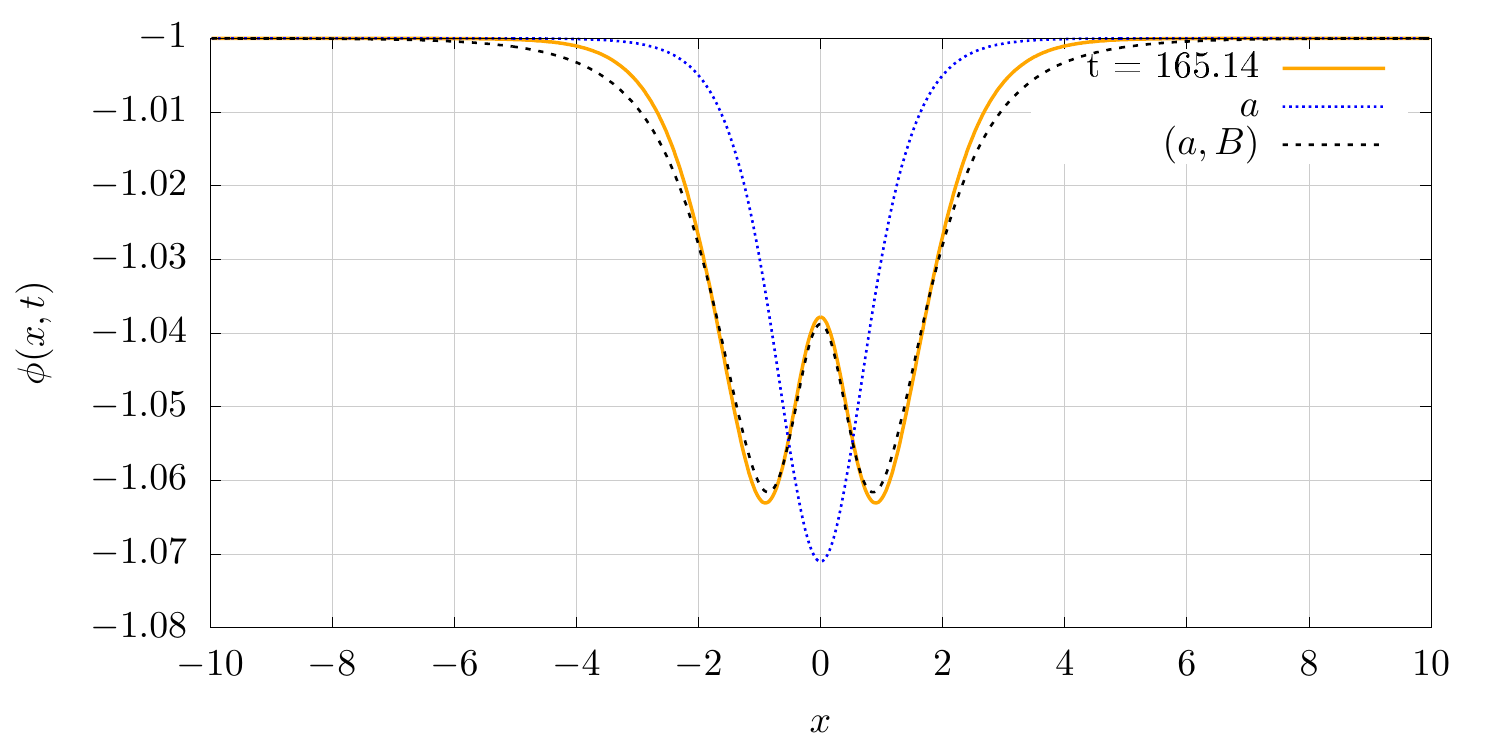} \includegraphics[width=0.49\textwidth]{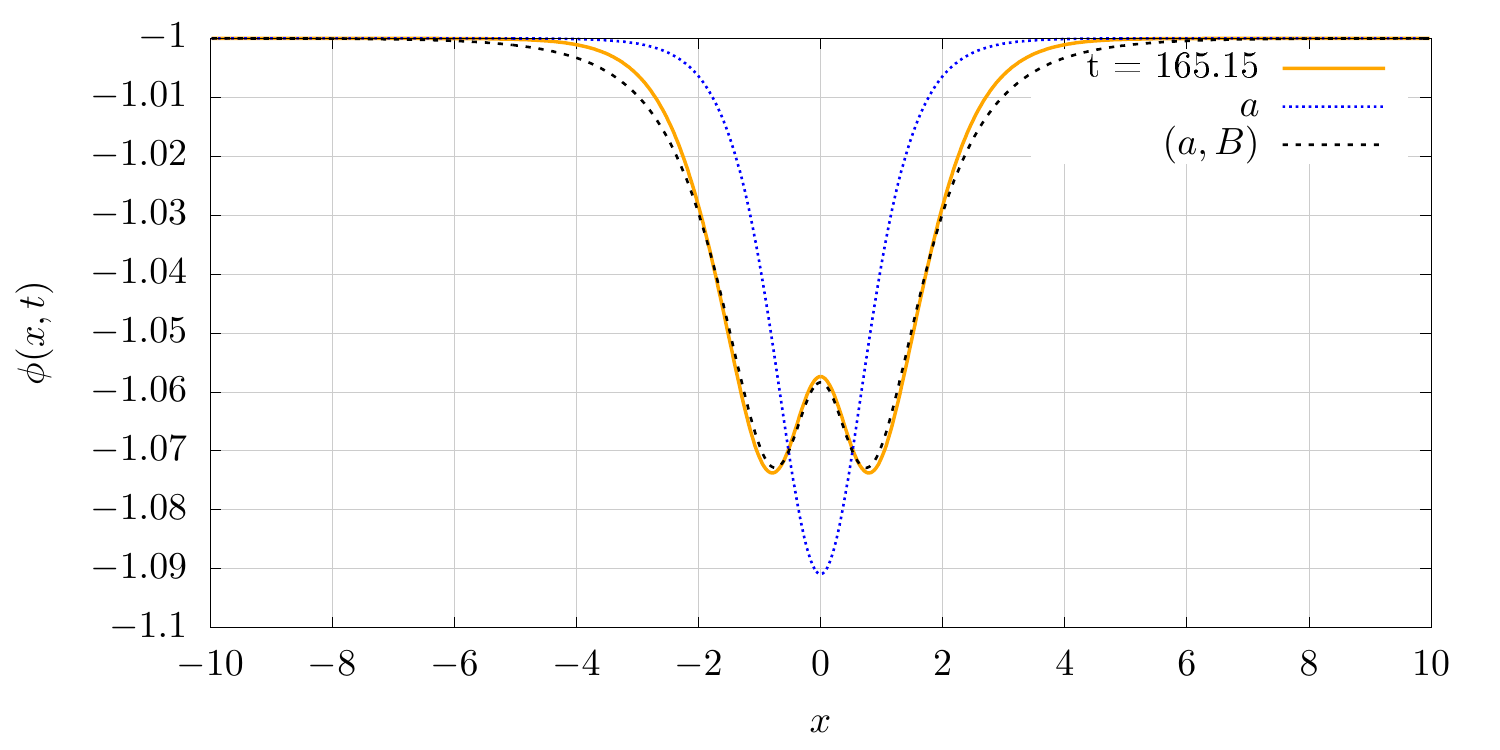}
\includegraphics[width=0.49\textwidth]{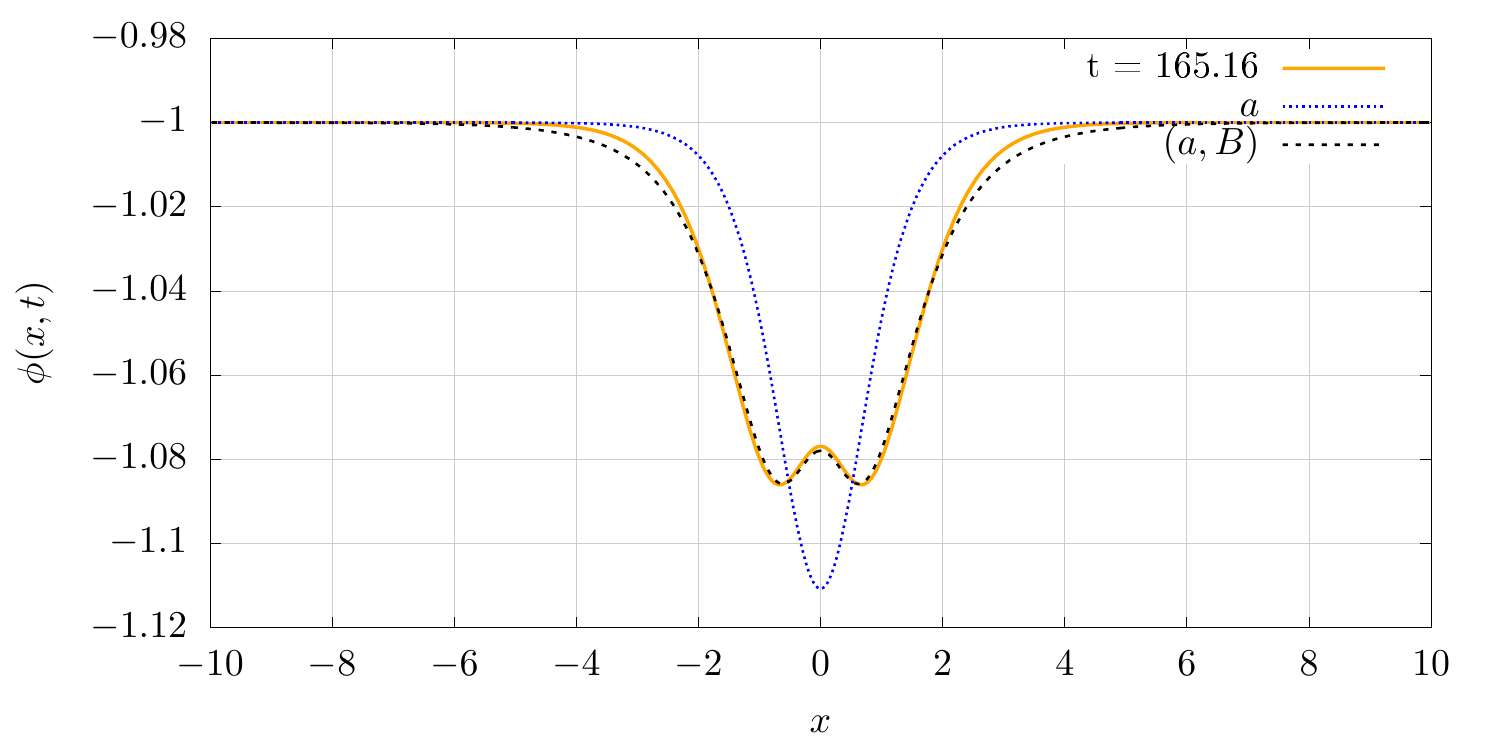} \includegraphics[width=0.49\textwidth]{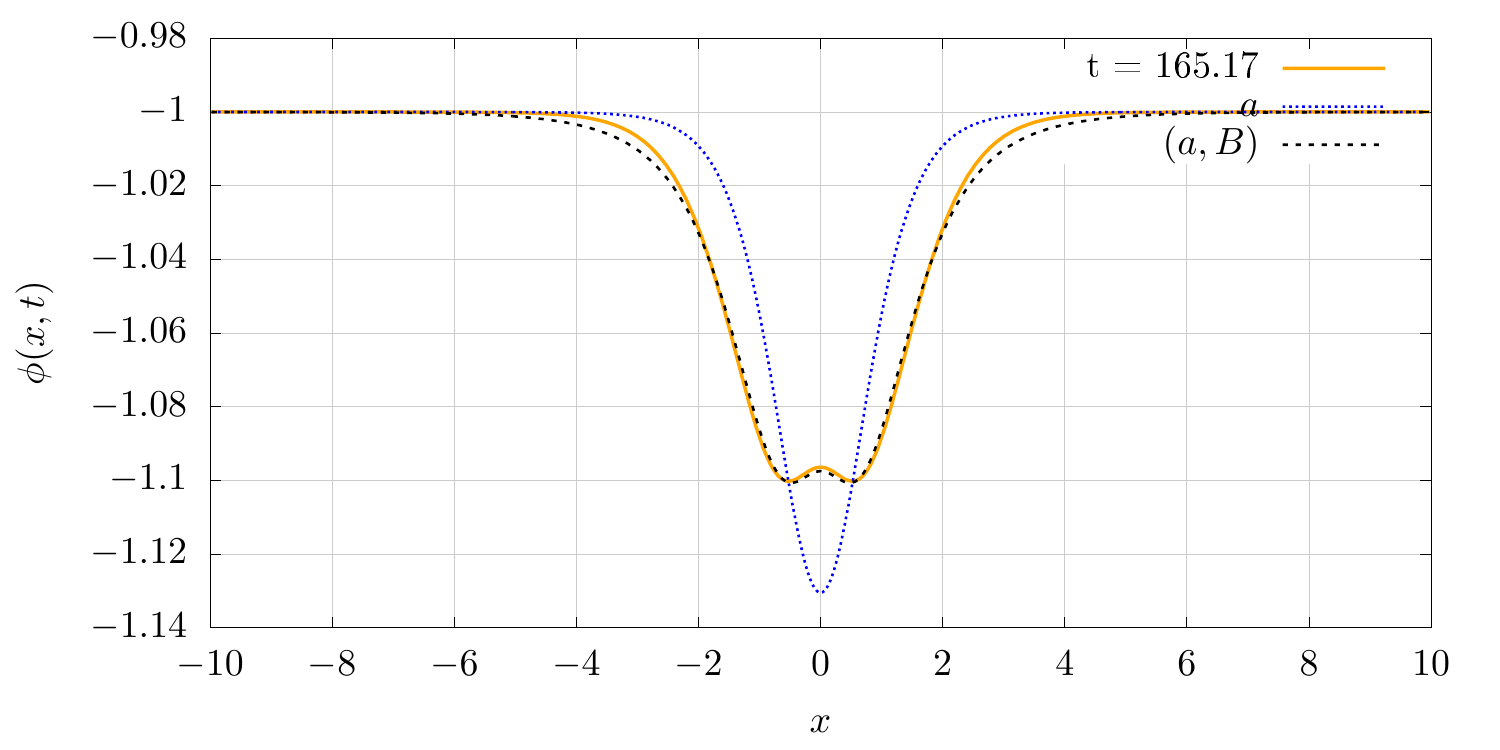}
\caption{Dynamical kink-antikink field configurations in $\phi^4$ theory 
(solid orange), with best fits to the 1-dimensional ($a$, fine dotted
blue) and 2-dimensional moduli space configurations ($(a,B)$, coarse
dotted black).}
\label{profiles-phi4}
\end{figure}

\begin{figure}
\includegraphics[width=0.49\textwidth]{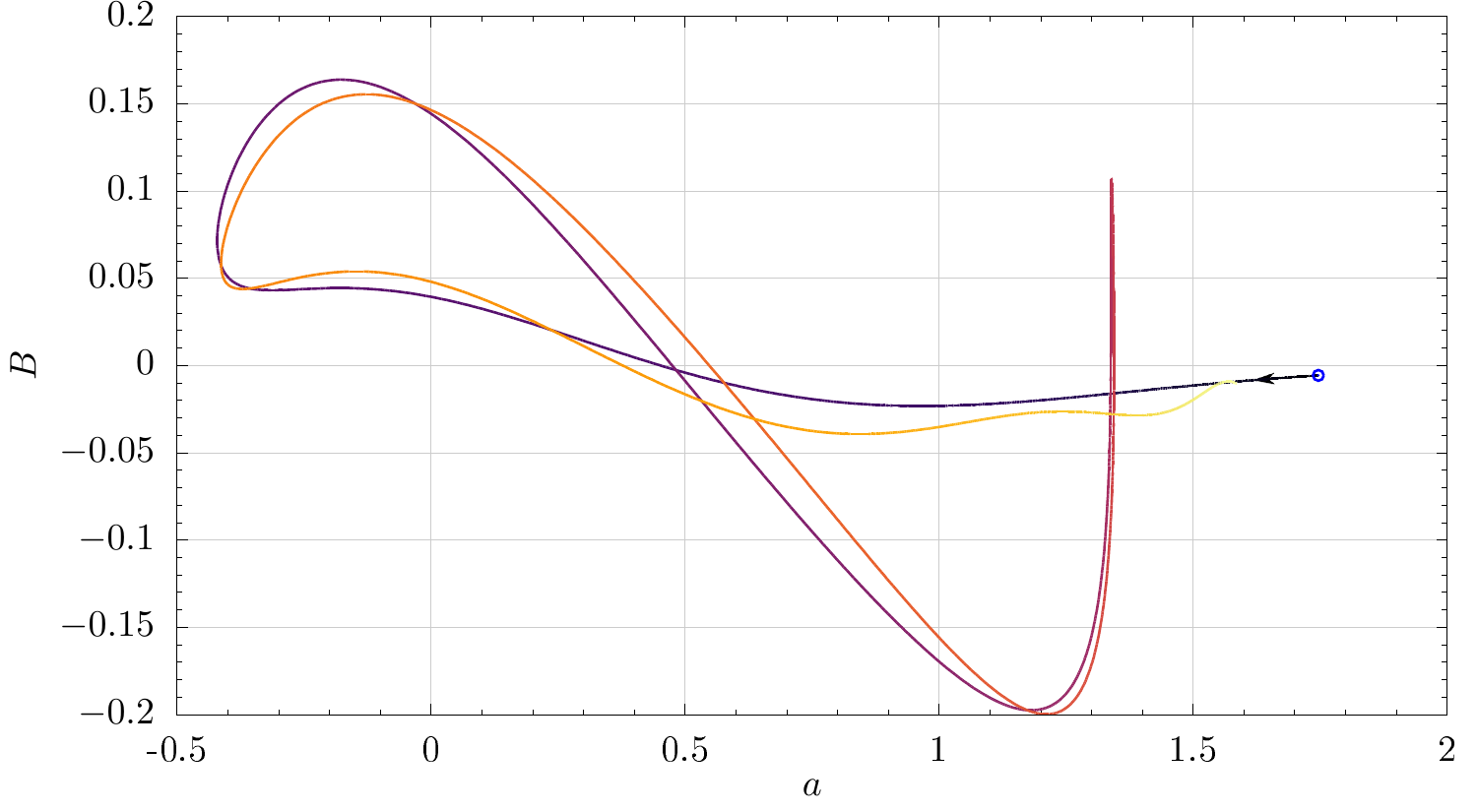} 
\includegraphics[width=0.49\textwidth]{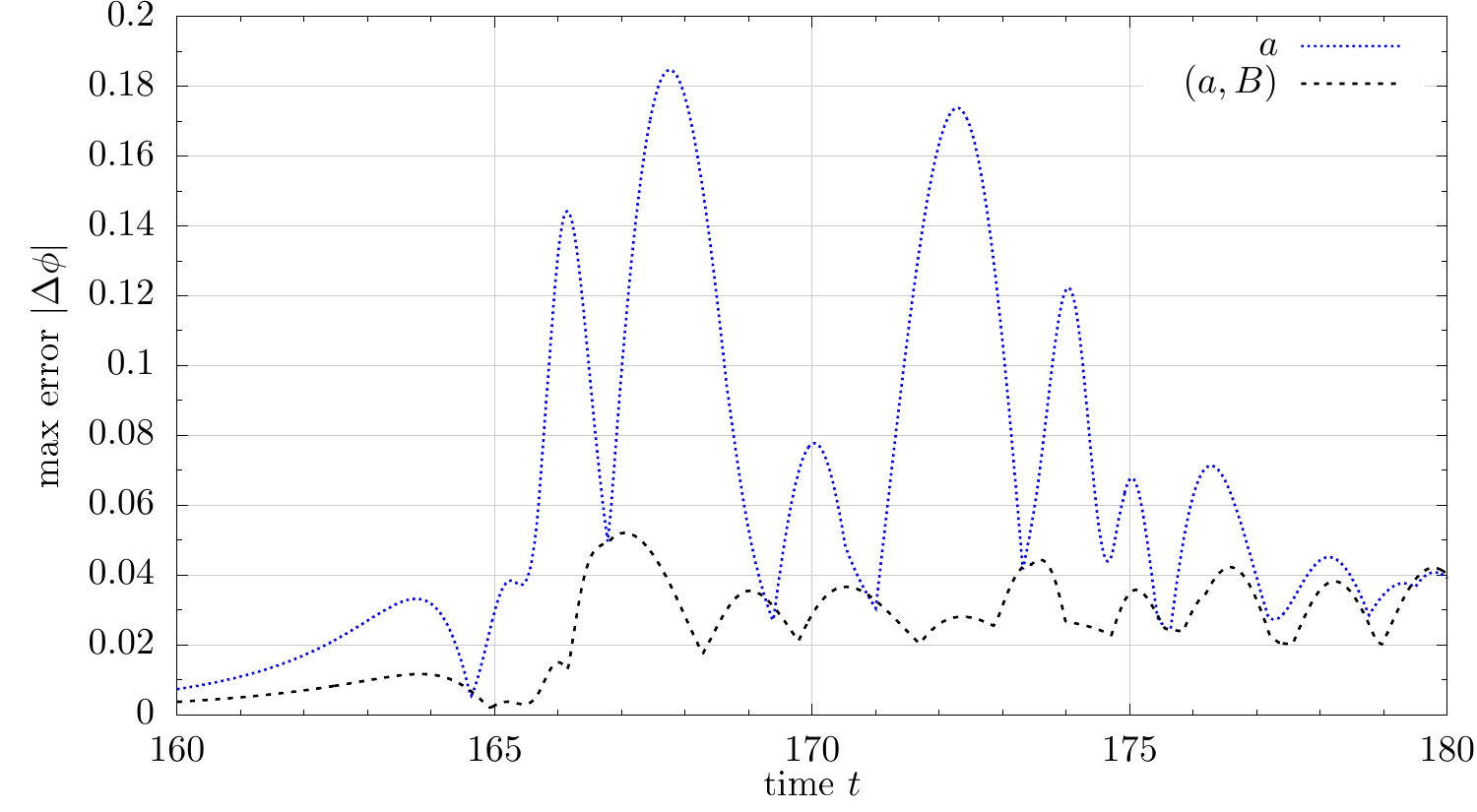} 
\caption{Kink-antikink scattering in $\phi^4$ theory. {\it Left:}
Phase plot for the fitted moduli $(a,B)$. The arrow indicates the 
initial point and colours denote passing time. {\it Right:} Maximal
field difference $|\Delta\phi|$ between the true dynamical field 
$\phi(x,t)$ and the closest moduli space configuration.}
\label{phase}
\end{figure}

The metric on the Sugiyama moduli space has been calculated by Pereira
et al. \cite{PLTC}, using contour integration to evaluate the 
integrals. It should not be difficult to recalculate the metric on 
the 2-dimensional subspace of configurations \eqref{collective-aB}, 
using the coordinates $a$ and $B$. The spatial integrals 
defining the metric coefficients are not affected, but need to 
be combined differently. Including distinct amplitudes for the two 
shape modes adds no further difficulties, because there is no additional
null-vector when the shape modes centred at $-a$ and $a$ are added
rather than subtracted.

We have found that kink-antikink dynamics in $\phi^4$ field theory follows
a trajectory through this 2-dimensional moduli space quite
well. We have evolved a kink and antikink starting nearly at rest at a large
separation, with initial velocity $v=0.05$, and have matched the field
configuration at each time $t$ to the configuration in the
moduli space with the smallest maximal difference in
field value $\phi$, thereby determining a trajectory $(a(t), B(t))$ in
moduli space. The true dynamical configuration and the moduli space 
configuration are shown in Fig.  \ref{profiles-phi4} at various times 
close to the first collision. Also shown is the closest 
configuration in the naive moduli space with the single
modulus $a$ -- the fit is clearly less good. The $(a,B)$ 
trajectory is plotted in Fig. \ref{phase}, left panel, for $0 \le t
\le 200$. It is smooth even though $a$ passes through zero a couple of
times, because of the use of the new coordinate $B$. The maximal field
difference is shown in Fig. \ref{phase}, right panel. It is very small
for all $t$ provided the 2-dimensional moduli space is used. What 
remains to be investigated is whether this trajectory is 
approximately reproduced by the moduli space dynamics for 
$a$ and $B$. Calculating the metric and potential on the 
2-dimensional moduli space (and also their derivatives) and solving 
for the dynamics is challenging and will be reported elsewhere.

\subsection{Weighted kink-antikink configurations} 
\label{sec-weighted-kak}
	
Here we discuss a variant of the naive kink-antikink
superposition, where the kink and antikink profiles are given a weight
that depends on their separation. The weight allows for
the kink to be modified by the nearby presence of the
antikink, and vice versa. The field configurations now have the
form
\be
\phi(x;a) = \tanh(a)(\tanh(x+a) - \tanh(x-a)) - 1 \,.
\label{KK*weight}
\ee
We do not include the shape modes here, although their inclusion would be
straightforward. The choice of weight function $\tanh(a)$ is motivated
by the identity
\be
\tanh(a)(\tanh(x+a) - \tanh(x-a)) - 1 = 
\frac{c - \cosh^2(x)}{c + \cosh^2(x)} \,,
\label{acidentity}
\ee
where $c = \sinh^2(a)$; conveniently, the expression on the right hand 
side occurs in the iterated kink scheme \cite{MOW} that we review in the next 
section. The identity is verified by writing
\be
\tanh(x+a) - \tanh(x-a) = \frac{\sinh(x+a)\cosh(x-a) -
\cosh(x+a)\sinh(x-a)}{\cosh(x+a)\cosh(x-a)} \,,
\ee
and then using hyperbolic analogues of trigonometrical addition and double
angle formulae.

For large positive $a$, the weighted superposition agrees with the
naive superposition. Curiously, it also agrees for
small positive $a$, because to leading order the weighted superposition
is
$\phi = \frac{2a^2}{\cosh^2(x)} - 1$ whereas the naive superposition is
$\phi = \frac{2a}{\cosh^2(x)} - 1$. The field configurations have the
same form, with a positive bump, but the parameter $a$ has changed 
its meaning. For intermediate $a$, the configurations are slightly
different, no matter how the parameters are matched.
 
However, the moduli spaces of naive and weighted superpositions 
are metrically quite
different as $a$ approaches zero, and if one attempts to extend beyond 
here. Recall that the naive superposition can be extended smoothly to
negative $a$, and the positive bump becomes a negative bump. On the
other hand, the weighted superposition (\ref{KK*weight}) is
symmetric in $a$, so its moduli space is completely covered by $a$ in
the range $0 \le a < \infty$. This moduli space has a boundary 
at $a=0$ and is geodesically incomplete, because $\phi$ 
depends quadratically on $a$ for small $a$, so
\be
\frac{\pr \phi}{\pr a} = O(a)
\ee
and the metric coefficient is $g(a) = O(a^2)$. $a$ is not a
good coordinate near $a=0$ for the weighted superposition, but 
$c = \sinh^2(a)$ is a better coordinate, as $c \approx a^2$ for small 
$a$, and in terms of $c$ the metric coefficient is $g(c) = O(1)$. This 
can be verified directly using the expression on the right hand 
side of (\ref{acidentity}). A motion in which $a$ evolves smoothly 
from positive through zero to negative is not acceptable
dynamically. (It is like the motion of a particle along a line that
stops and reverses, even though no force acts.) Instead, a motion 
where $c$ evolves smoothly through zero is acceptable.

The natural range of $c$ is the interval $(-1,\infty)$, as $\phi$ has a
singularity at $x=0$ when $c = -1$. When $c < 0$, the
field configuration has a negative bump with a different shape from
the bump that occurs for negative $a$ in the naive kink-antikink
superposition. The identity (\ref{acidentity}) is
still valid for $c < 0$ if suitably interpreted. $a$ becomes imaginary,
so let us write $a = i{\tilde a}$ with ${\tilde a}$ positive. (We will
comment further on the signs of $a$ and ${\tilde a}$ below.) Recall the
relations
\bea
\sinh(i{\tilde a}) = i\sin({\tilde a}) \,, &&  \sin(i{\tilde a}) =
i\sinh({\tilde a}) \,, \nonumber \\
\cosh(i{\tilde a}) = \cos({\tilde a}) \,, &&  \cos(i{\tilde a}) =
\cosh({\tilde a}) \,.
\eea
Therefore, $c = -\sin^2({\tilde a})$ for negative $c$, and we can
express the weighted superposition of kink and antikink
(\ref{KK*weight}) as
\be
\phi(x;{\tilde a}) = i\tan({\tilde a}) (\tanh(x+i{\tilde a}) - \tanh(x -
i{\tilde a})) - 1 \,.
\label{cxkinks}
\ee
This expression is real, and symmetric in ${\tilde a}$. Use of
the subtraction formula for the $\tanh$ function leads back to 
the formula for $\phi$ in terms of $c$.

The interpretation of (\ref{cxkinks}) is that the kink and antikink have
imaginary positions $\pm i{\tilde a}$. The moduli space is well defined
in terms of $c$ and motion through $c=0$ is unproblematic, but 
it corresponds to a 90-degree
scattering of the positions of the kink and antikink. Such scattering is
familiar in the dynamics of solitons in higher dimensions, for example
for vortices and monopoles, where the scattering occurs
in a real 2-dimensional plane. Here, remarkably, the scattering occurs
in the complex plane of the 1-dimensional kink position parameter $a$.
This type of scattering is only possible because the sign of $a$ is not
fixed. The kink and antikink are initially located on the real axis at
$-a$ and $a$, but because of the weight factor, there is no effect if
$a$ is replaced by $-a$. The kink and antikink later appear on the
imaginary axis at ${\tilde a}$ and $-{\tilde a}$, and note that because
the amplitude $i\tan({\tilde a})$ is imaginary, the kink and antikink
cannot now be distinguished. Therefore
the scattering does not break the reflection symmetries of the
complexified $a$-plane. A pair of points scatters from the real to
the imaginary axis, in the same way that the algebraic roots of $z^2 = c$
scatter as $c$ passes through zero along the real axis.

The moduli space with coordinate $c$ is geodesically complete, for $c$
extending to $-1$. This could be verified by calculating the metric
factor $g(c)$, but is easier to see as follows. The limiting
(singular) field configuration is
\be
\phi(x;c=-1) = -1 - \frac{2}{\sinh^2(x)} \,,
\ee
whereas $\phi(x;c=0) = -1$. In the field configuration space, the 
squared distance (\ref{s2}) between these configurations 
is the integral of $\frac{4}{\sinh^4(x)}$, but this is divergent. 
The distance in the moduli space between $c=0$ and $c=-1$ (which is 
not a straight path in field configuration space, so not shorter) 
is therefore also infinite.

In summary, the moduli space of a superposed weighted kink and antikink
is an interesting alternative to the more familiar moduli space of the 
naive superposition. For this new moduli space to be geodesically 
complete, one must allow for the kink and antikink to scatter in 
the complexified plane of their positions $a$ and $-a$. However the 
field remains real, and in terms of the modulus
$c$ the field just transitions from having a positive bump to a negative
bump as $c$ passes through zero. Further investigation is needed to see
if this moduli space is better than the moduli space of the naive
superposition for modelling kink-antikink dynamics in the
full field theory. It is almost certainly necessary to include shape
modes in either case.

\section{Iterated Kinks}

In this section we briefly review the iterated kink equation that was
introduced by three of the present authors in \cite{MOW}. It is a
sequence of ODEs,
\be
\frac{d \phi_n}{dx} = -(1 - \phi_n^2)\phi_{n-1} \,, \quad n=1,2,3,\dots
\,,
\ee
where we fix $\phi_0(x) = -1$. For all $n$ we impose the boundary
condition $\phi_n(x) \to -1$ as $x \to -\infty$.
The equations are solved sequentially, and one can stop at any $n$. We
will only discuss the solutions in any detail up to $n=3$. For odd (even) 
$n$, the generic solution $\phi_n$ approaches $+1$
($-1$) as $x \to \infty$. For odd $n$, we therefore impose the boundary
condition $\phi_n(x) \to +1$ as $x \to \infty$; then the solution 
always lies between $-1$ and $+1$. For even $n$, the solutions are 
essentially of two types: a solution either lies between $-1$ and 
$+1$, or is everywhere less than $-1$. These types are separated 
by the constant solution $\phi_n(x) = -1$. As each equation of 
the iterated kink sequence is of first order, its solution has one 
constant of integration. If we retain all of these, the $n$th iterate 
$\phi_n$ depends on $n$ arbitrary constants -- the moduli of the 
$n$th iterate. For even $n$, the constant solution $\phi_n(x) = -1$ 
lies in the interior of the moduli space.

The iterated kink equation is a systematic extension of the idea of a
kink equation with impurity $\chi(x)$ \cite{BPS-imp, BPS-susy, Solv}
\be
\frac{d \phi}{dx} = -(1 - \phi^2)\chi(x) \,,
\ee
which in turn is a generalisation of the basic first order ODE for
a $\phi^4$ kink or antikink, $\frac{d \phi}{dx} = \pm(1 - \phi^2)$.

The iterated kink solutions are interesting because a
large part of the moduli space of the $n$th iterate consists of field
configurations with $n$ alternating kinks and antikinks at arbitrary
separations (starting with a kink on the left). Other parts of the 
moduli space consist of configurations
where a kink-antikink pair, or more than one pair, have approached each
other to form a bump. The moduli space of the $n$th
iterate therefore usefully describes configurations of $n$ kinks
and antikinks. However these configurations do not incorporate the 
standard shape mode deformations of a kink or antikink.

We conjecture that each of these moduli spaces has a positive definite
metric and is metrically smooth and geodesically complete, almost 
everywhere. We have not proved this, 
but it is clearly true for $n=1$ and $n=2$, and the numerical study
for $n=3$ and $n=4$ reported in \cite{MOW} indicates that the partial 
derivatives of the field $\phi_n(x)$ with respect to all moduli are 
almost everywhere non-zero and linearly independent. The exceptional
points in moduli space are the constant configurations $\phi_n(x) =
-1$, for even $n$, 
where a rigid translation has no effect. We will focus on the relative 
motion of kinks, so this problem with translations does not occur.

Let us now recall the first few iterates. For $n=1$, we have
the standard $\phi^4$ kink equation, with solution $\phi_1(x;a) =
\tanh(x-a)$. The moduli space is the real line
with its Euclidean metric. For $n=2$, the field $\phi_1$ acts as an 
impurity. As $a$ shifts, the solution $\phi_2$ simply
shifts with it. Let us therefore ignore this translational modulus, 
and assume that $\phi_1(x) = \tanh(x)$. Then one expression for 
the solutions $\phi_2$ is
\be
\phi_2(x;c) = \frac{c - \cosh^2(x)}{c + \cosh^2(x)} \,.
\label{iterphi2}
\ee
These configurations, which appeared in Section 4, are non-singular
for $-1 < c < \infty$, and are illustrated in ref.\cite{MOW}. 
Note that for $c>0$, $\phi_2$ is 
between $-1$ and $+1$; for $c=0$, $\phi_2(x) = -1$; and for $-1 < c < 0$, 
$\phi_2$ is everywhere less than $-1$. The moduli space 
has no metric singularity at $c=0$.

The third iterate $\phi_3(x)$ is obtained from the second
iterate $\phi_2$ by integrating the relevant ODE. We will only consider 
the solutions that are antisymmetric in $x$, although
the general solution with broken reflection symmetry is known and
illustrated in \cite{MOW}. The only modulus is then $c$, arising from
$\phi_2$. The antisymmetric solutions $\phi_3(x;c)$ are kink-antikink-kink
configurations when $c$ is large, having an antikink at the origin and
kinks at equal separation on either side. As $c$ approaches zero and
becomes negative, the kinks approach and annihilate the antikink,
leaving a single kink modified by a variant of the shape mode. In the next
section we look at the moduli space of these configurations in more
detail.

\section{Kink-Antikink-Kink Moduli Spaces in $\phi^4$ Theory}

Here we investigate three candidate moduli spaces for kink-antikink-kink 
configurations in $\phi^4$ theory, antisymmetric in $x$. Each moduli
space is 1-dimensional.

\subsection{Naive superposition of kink, antikink and kink}

An obvious choice for a kink-antikink-kink configuration is the
generalisation of the naive kink-antikink superposition,
\be
\phi(x;b) = \tanh(x+b) - \tanh(x) + \tanh(x-b) \,.
\label{KK*Knaive}
\ee
The modulus $b$ parametrises the positions of the two kinks, and the
antikink is at the origin. $\phi(x;b)$ is symmetric in 
$b$. For positive $b$ we have an equally spaced 
chain of kink, antikink and kink. The kinks approach the antikink as 
$b \to 0$, and at $b=0$ merge to form the standard single kink 
at the origin. For negative $b$ this process reverses, so the 
moduli space is really just the half-line with $b$ non-negative.
  
The metric on the moduli space depends on the derivative with 
respect to $b$,
\be
\frac{\partial \phi }{\partial b} (x;b) = 
\left( \frac{1}{\cosh^2(x+b)} -\frac{1}{\cosh^2(x-b)} \right)\,.
\ee
This is antisymmetric in $b$ and vanishes at $b=0$. The resulting metric 
is
\be
g(b)=\frac{2}{3} \, \frac{ -24 b \cosh (2b) + 9 \sinh (2b) + \sinh (6b)}
{\sinh^3(2b)} \,, 
\label{naive-kakk-g}
\ee
and also vanishes at $b=0$. The metric on the half-line 
is therefore not complete and $b$ is not globally a good coordinate. 
However, this problem is resolved by extending the moduli space.

Usefully, the field superposition \eqref{KK*Knaive} 
is (minus) the product of the iterated configuration 
$\phi_2$, given by \eqref{iterphi2}, and a kink at the origin 
$\phi_1(x;0) = \tanh(x)$,
\be
\phi(x;b) = 
- \left( \frac{\sinh^2(b) - \cosh^2(x)}{\sinh^2(b) + \cosh^2(x)}
\right) \tanh(x) = - \phi_2(x;c) \phi_1(x;0) \,, 
\label{naive-iter}
\ee
where $c=\sinh^2(b)$ and is therefore non-negative for real $b$. 
$c$ is a better choice for the modulus, as we can straightforwardly 
extend the moduli space to negative values of $c$, although this requires
$b$ to be extended to imaginary values $b = i\tilde{b}$, with
$\tilde{b}$ positive. Then $c = -\sin^2(\tilde{b})$, so the range of
$\tilde{b}$ is $0 \le \tilde{b} < \half\pi$. The new field
configurations are more compressed than the standard kink at the origin. 
$\phi(x;{\tilde b})$ becomes not only steeper as $\tilde{b}$ grows, but 
also acquires negative and positive bumps outside the range
of field values $[-1,1]$ as $\tilde{b}$ approaches $\half\pi$. These
bumps are produced by the large negative bump in $\phi_2$ for $c$
approaching $-1$.
\begin{figure}
\includegraphics[width=0.49\textwidth]{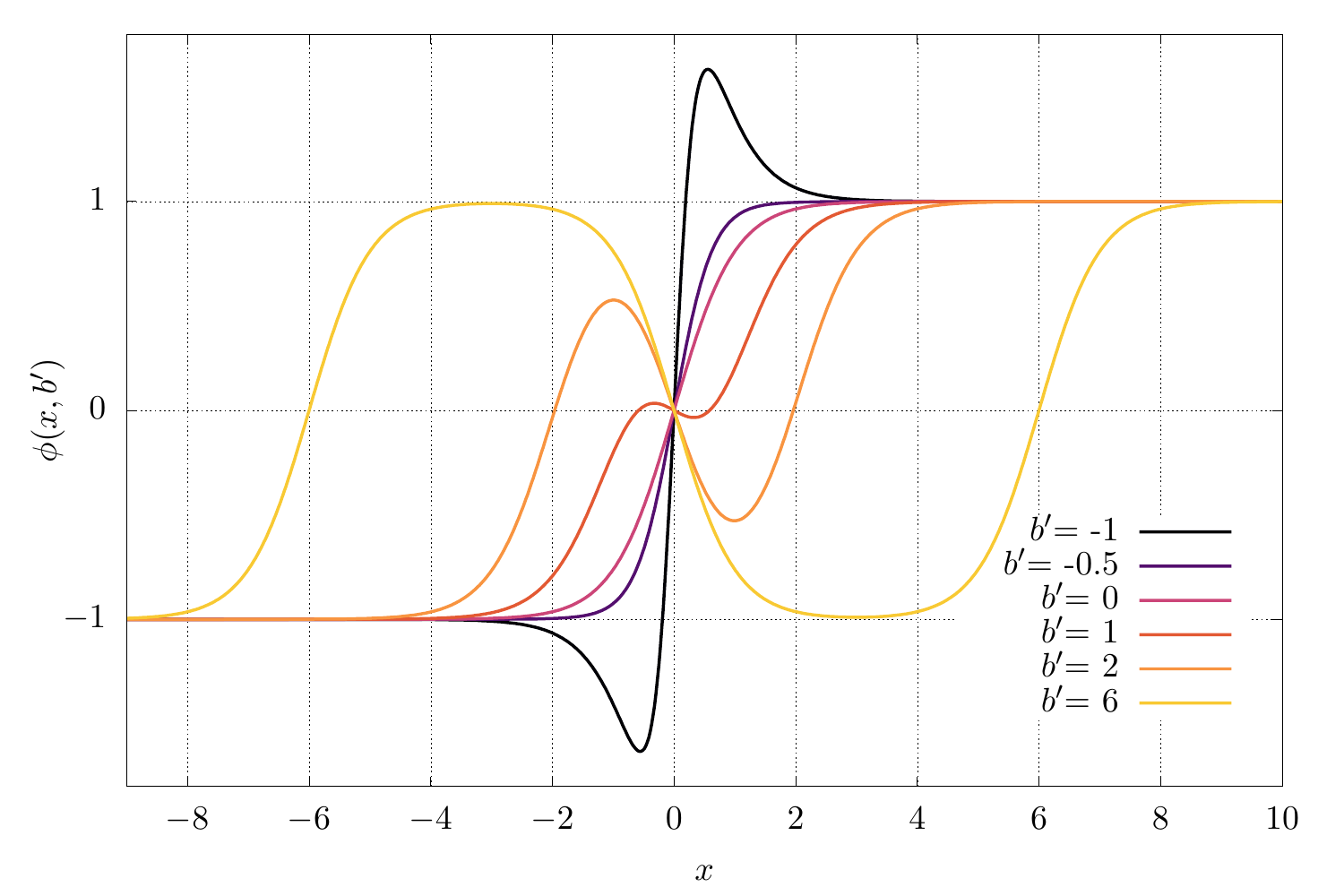}
\includegraphics[width=0.49\textwidth]{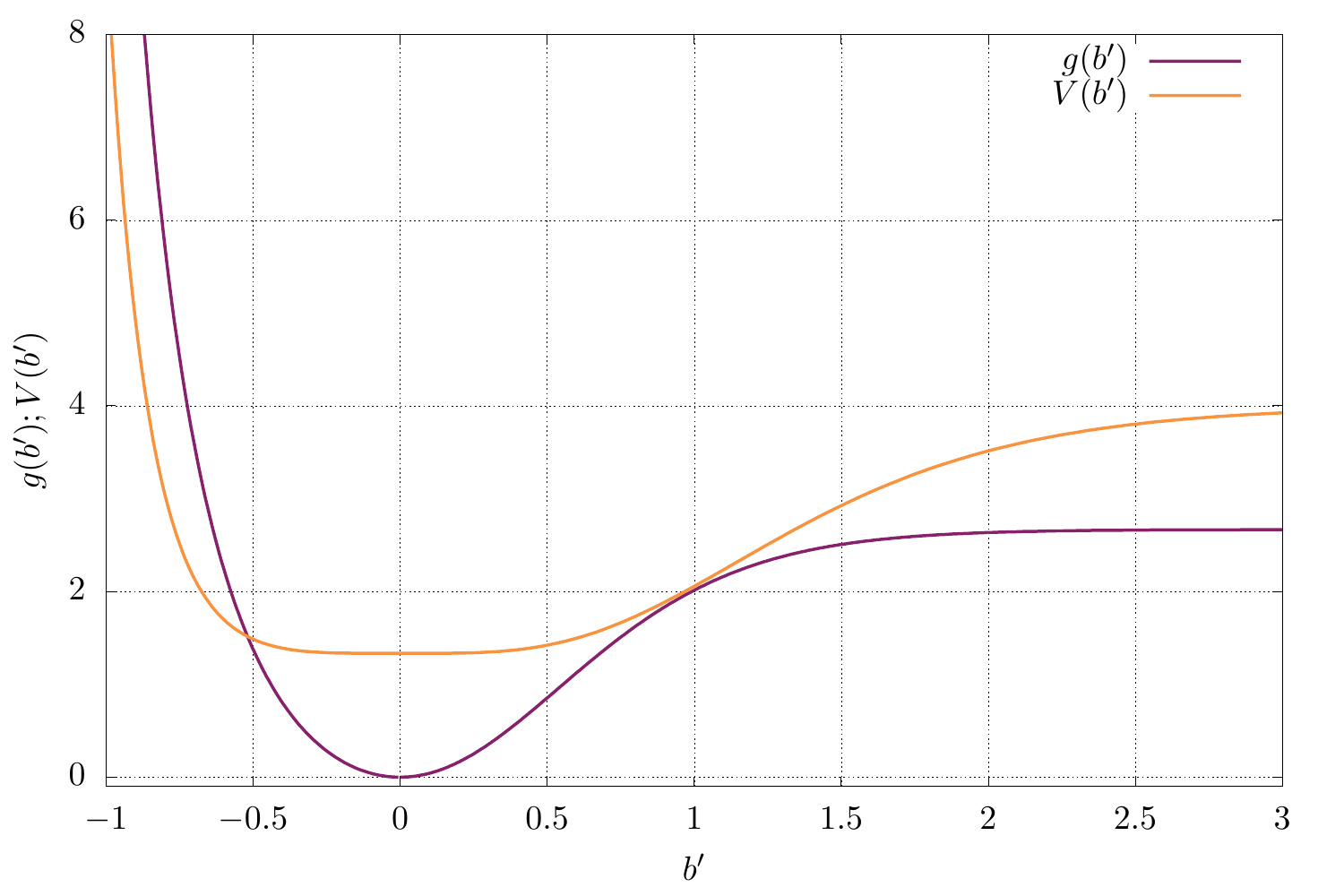}
\caption{Naive superposition of kink, antikink and kink. {\it Left:}
Profiles for several values of the extended modulus $b'$. {\it Right:}
The moduli space metric and potential.}
\label{3-naive-plot}
\end{figure}

At $c = -1$ the potential energy is
infinite, so in a dynamical kink-antikink-kink motion $c$
decreases, stops before it reaches $-1$, and increases again. 
In the complex plane of the modulus $b$, there is 90-degree 
scattering during a kink-antikink-kink collision, which
occurs in reverse as the motion reverses. From the
perspective of the naive superposition, the kinks scatter but the
antikink remains at rest at the origin. In Fig. \ref{3-naive-plot} 
we plot the field configurations as well as the metric and potential
on moduli space as a function of $b$. We show these
for the real and imaginary ranges of $b$ simultaneously
by introducing a new real coordinate $b'$ defined by
\be
b' = \left\{
\begin{array}{cl}
b &  \text{for} \;\; b \in [0,\infty) \\
-\tilde{b} & \text{for} \;\; \tilde{b} \in  \left[0, \half\pi \right) \,.
\end{array}
\right. 
\label{b'}
\ee
 
Note that, under the replacement $b \to i \tilde{b}$, the metric
(\ref{naive-kakk-g}) acquires an additional minus sign to compensate 
$\dot{b}^2$ being negative in the kinetic part of the Lagrangian. 
There is not a genuine sign problem, and the metric is positive if 
$c$ is used as the coordinate.
  
\subsection{Iterated kink-antikink-kink}

\begin{figure}
\includegraphics[width=0.49\textwidth]{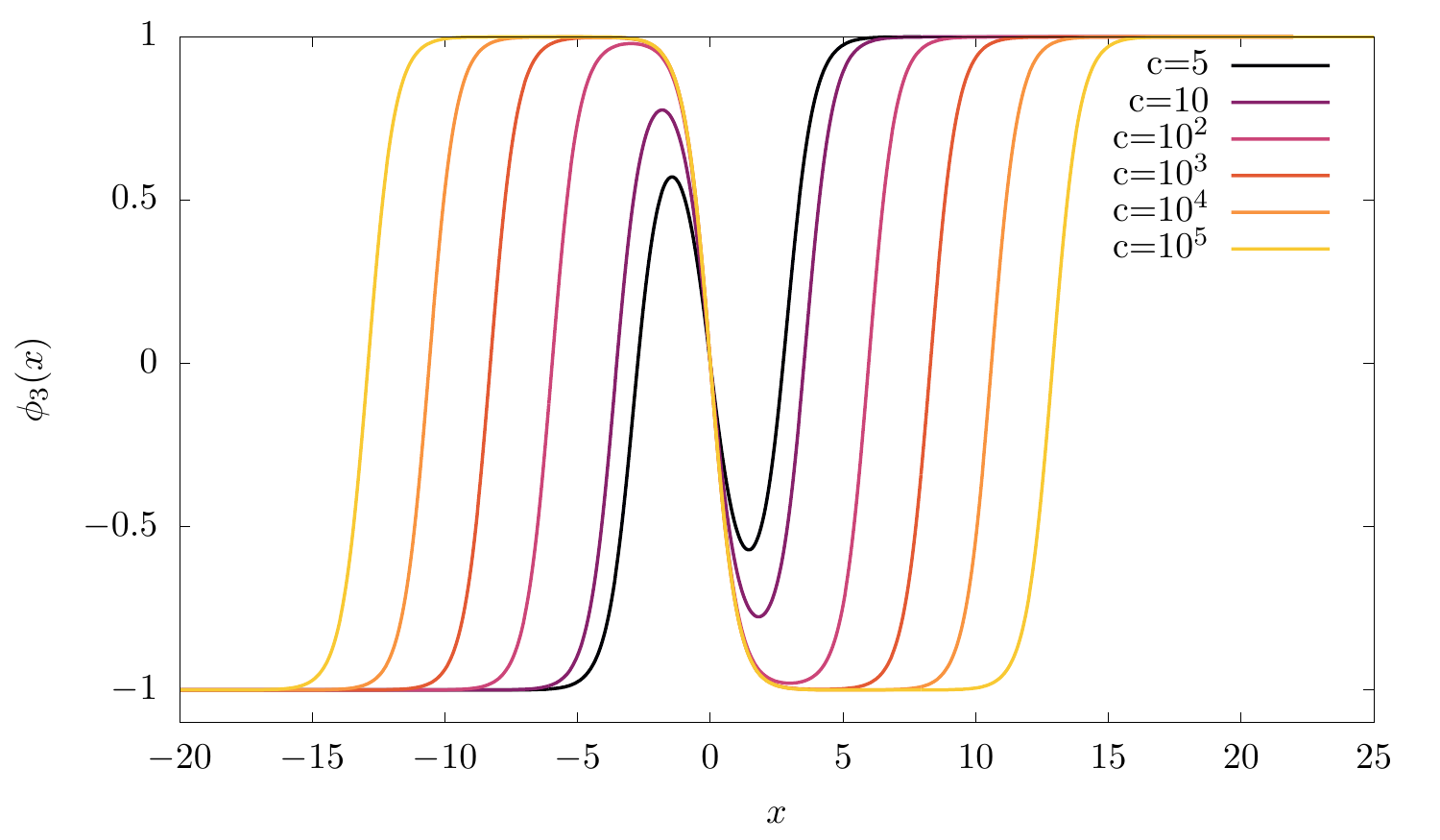}
\includegraphics[width=0.49\textwidth]{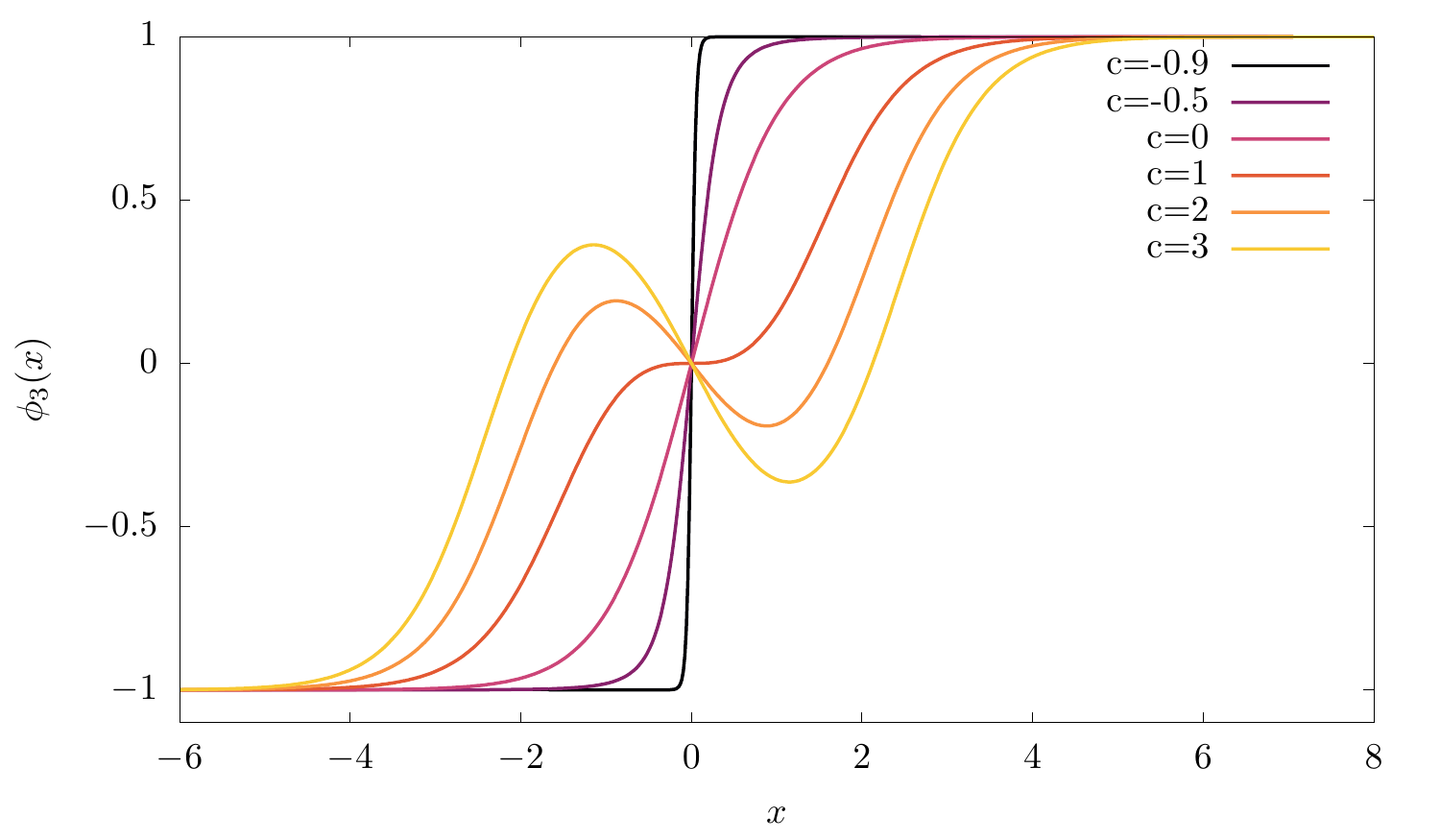}
\caption{Iterated $\phi_3$ solution for several values of the modulus $c$.}
\label{phi3-plot}
\end{figure}

The iterated kink scheme gives an alternative moduli space of
kink-antikink-kink configurations. Iterating the $\phi_2$ solution, with
its modulus $c$, we obtain the $\phi_3$ solution 
\be
\phi_3(x;c)=\tanh \left(x - 2\sqrt{\frac{c}{1+c}} 
\mbox{ arctanh} \left( \sqrt{\frac{c}{1+c}} \tanh (x) \right)
\right) \,. 
\label{phi3-exact}
\ee
There is no further modulus derived from the constant of integration
because we select antisymmetric configurations with equal
kink-antikink-kink spacings. The formula \eqref{phi3-exact} is valid for all 
$c > -1$, even though the square root is imaginary if $c < 0$. 
$\phi_3$ is the standard kink for $c=0$, and is a compressed kink
for negative $c$, becoming infinitely steep as $c \to -1$. 
So $\phi_3$ is qualitatively similar to the
naive superposition with $b$ continued to imaginary values, but 
here the field value is always confined to $[-1,1]$, and does not have
bumps. These field configurations are shown in Fig. \ref{phi3-plot}.
\begin{figure}
\center \includegraphics[width=0.53\textwidth]{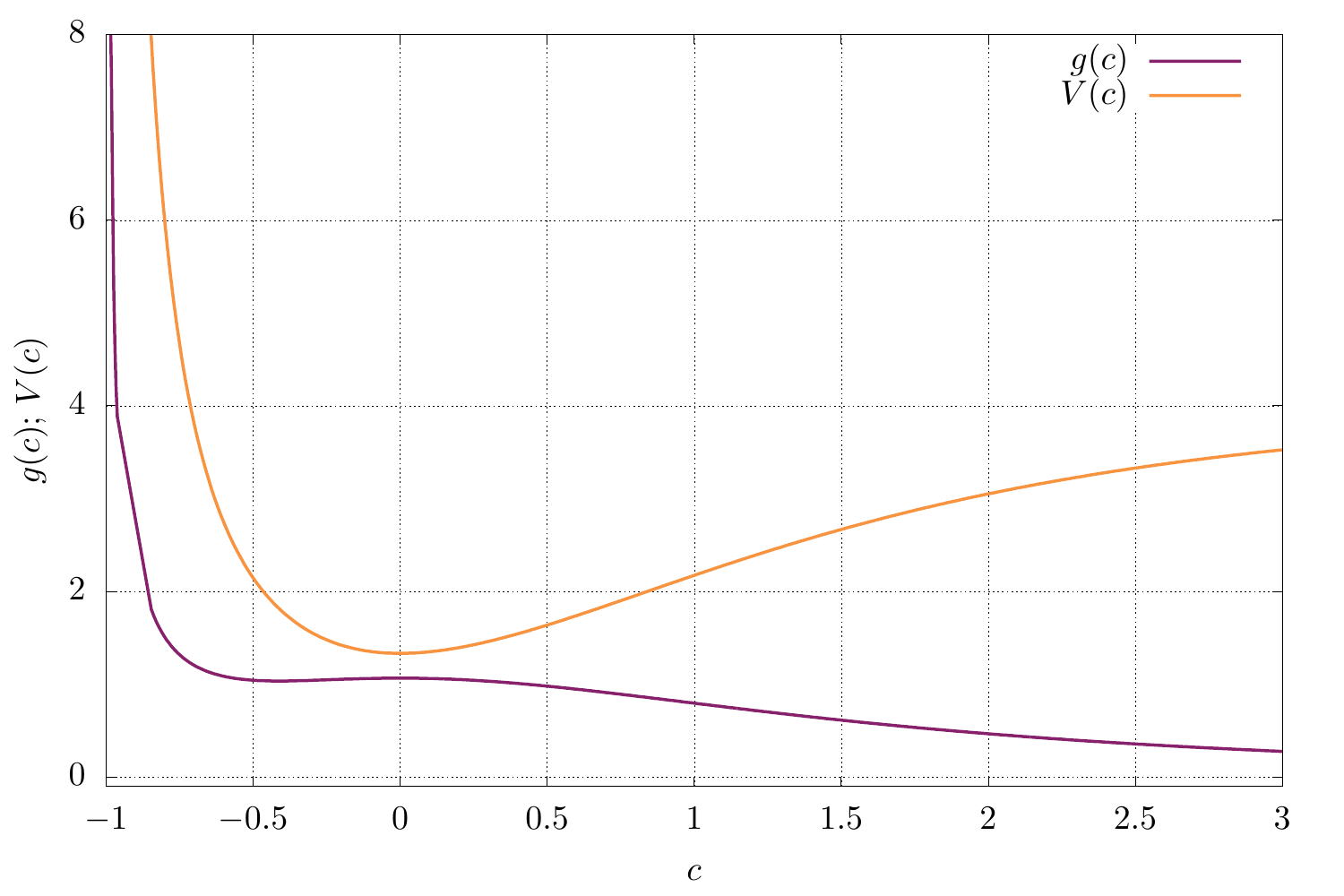}
\caption{Moduli space metric $g(c)$ and the potential $V(c)$ for the
iterated $\phi_3$ solutions.}
\label{phi3-metric-plot}
\end{figure}

The metric $g(c)$ on this moduli space, together with the potential 
$V(c)$, are shown in Fig. \ref{phi3-metric-plot}. $g(c)$ is positive
so $c$ is a good coordinate. In particular,
\be
\frac{\partial \phi_3 }{\partial c}(x;c=0) 
= -2 \frac{\sinh(x)}{\cosh^3(x)} \,,   
\ee
which gives $g(0)=\frac{16}{15}$. The moduli space is still
incomplete, with boundary at $c=-1$, but this does not matter as $V$ is 
infinite here. For example, the squared distance (\ref{s2}) between 
the standard kink $\phi_3(x; c=0)$ and the boundary configuration 
$\phi_3(x;c \to -1)$ is $s^2=2(-1+\ln 4)$.
  
\subsection{Weighted superposition of kink-antikink-kink}

Recall that the kink-antikink configuration $\phi_2$ can be
expressed exactly as a weighted superposition of kink and
antikink, so it could be interesting to find a moduli
space of weighted kink-antikink-kink configurations. A general
formula introducing a weight $\Lambda(b)$ is
\be
\phi(x;b) = \Lambda(b) \tanh(x+b) + (1 - 2 \Lambda(b)) \tanh(x)
+ \Lambda(b) \tanh(x-b) \,.
\label{weightedphi3}
\ee
This configuration is antisymmetic in $x$ and satisfies the vacuum 
boundary conditions. The naive superposition has $\Lambda = 1$, but 
allowing non-constant $\Lambda$ could be an improvement. 

\begin{figure}
\includegraphics[width=0.49\textwidth]{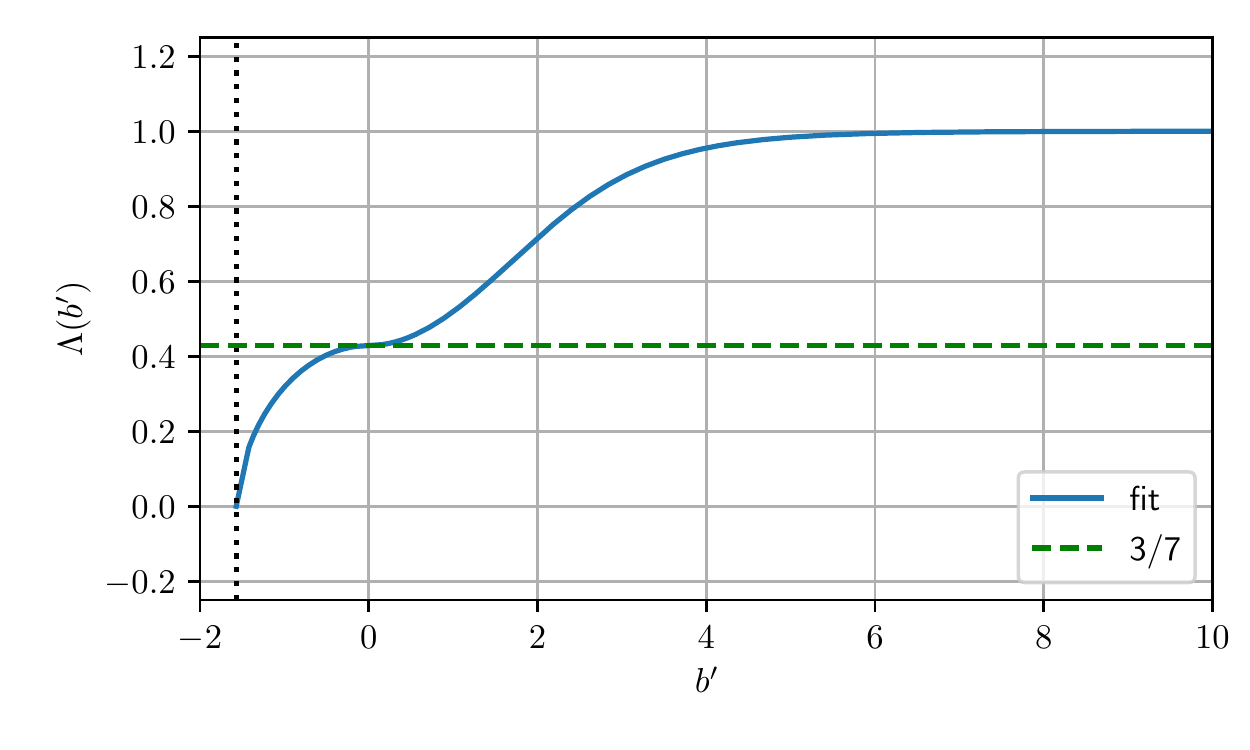}
\includegraphics[width=0.49\textwidth]{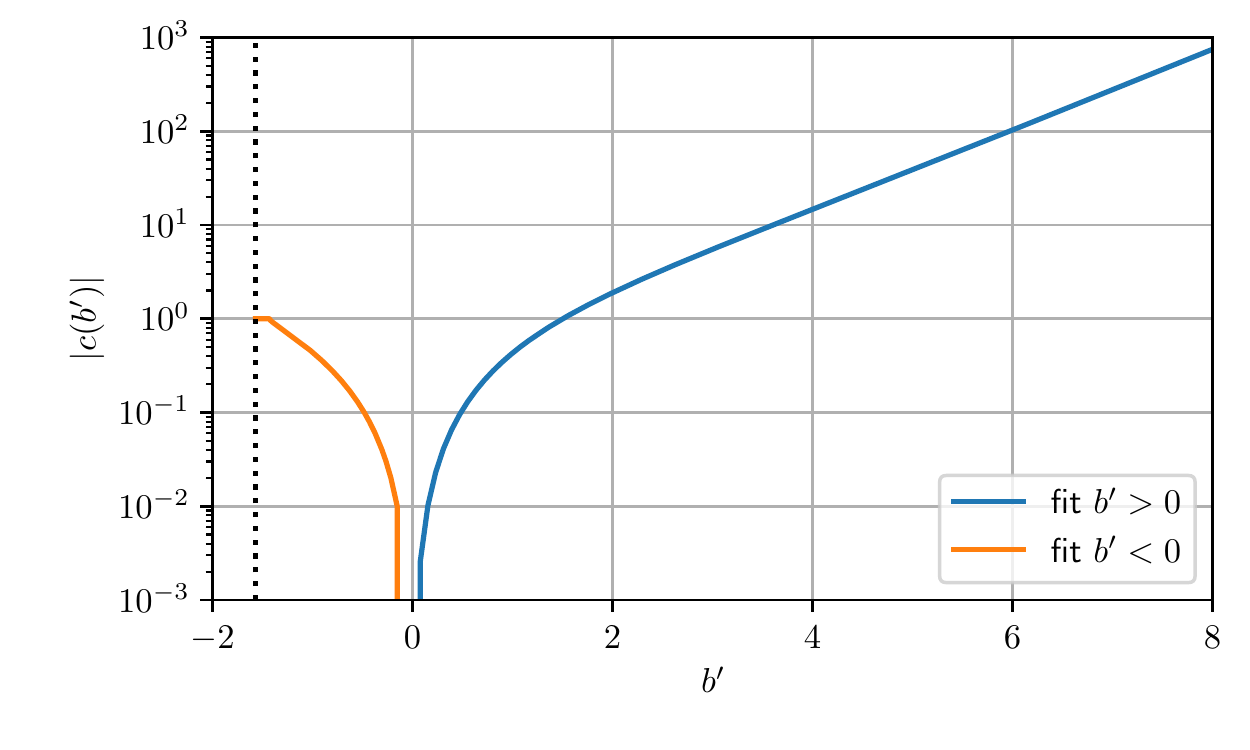}
\caption{{\it Left:} $\Lambda (b')$. {\it Right:}  $c(b')$.}
\label{Lambda}
\end{figure}

The weighted superposition is quite close to the iterated kink
solution $\phi_3(x;c)$ for a suitable choice of $b$ and $\Lambda(b)$
in terms of $c$. For $c$ negative, $b$ becomes imaginary and we 
again introduce the extended modulus $b'$.
In Fig. \ref{Lambda} we plot the optimal $b$ and $\Lambda(b)$,
found using a least squares fit of $\phi(x;b)$ to $\phi_3(x;c)$.
For large $c$, the kinks are well-separated from the antikink, so
$\Lambda \approx 1$ and $b \approx \log(4c)$. When $c$ is close to 
zero, the best fit is with $b \approx \sqrt{\frac{7c}{3}}$ 
and $\Lambda = \frac{3}{7}$. Strictly speaking, $\Lambda$ is 
indeterminate when $c=0$, but these values of $b$ and $\Lambda$ give 
the best fit nearby, giving an exact fit up to quadratic order in 
$c$ in the Taylor expansion of $\phi_3(x;c)$. $\Lambda$ decreases 
towards zero as $c$ approaches $-1$. Here $\phi_3$ approaches a 
step function, and the fit is not very good.

These weighted kink-antikink-kink superpositions with numerically 
determined coefficients are probably not very useful, especially since
they are close to the iterated kink configurations $\phi_3$. More interesting
could be a moduli space of configurations obtained using a suitable 
analytic formula for the weight $\Lambda(b)$. 

\section{Conclusions}

We have constructed a number of moduli spaces -- collective coordinate 
manifolds -- for field configurations of kink and antikink solitons in both 
$\phi^4$ theory and sine-Gordon (sG) theory. These are smooth, 
finite-dimensional submanifolds of the infinite-dimensional 
field configuration spaces. Only in parts of the moduli space do the 
moduli correspond to the real positions of the kinks and antikinks as
separated particles. The moduli have a different interpretation when a kink and
antikink annihilate and produce a bump in the field, or when a
kink-antikink-kink configuration turns into a compressed single kink.
For the $\phi^4$ kinks we have incorporated an amplitude for the 
shape mode oscillation (the discrete, 
normalisable kink oscillation mode whose frequency is below 
the continuum). For kink-antikink and also kink-antikink-kink 
configurations, we have assumed a spatial reflection symmetry 
to simplify the analysis.

By restricting the field theory Lagrangian to field 
configurations evolving through moduli space, we obtain, through
spatial integration, a reduced Lagrangian with 
kinetic and potential terms on moduli space, whose equation of motion 
is an ODE. The coefficient matrix of the kinetic term is 
interpreted as a Riemannian metric on moduli space. (In several of 
our examples there is just one modulus and the metric is a single 
function.)  A guiding principle is that this metric should be {\it either} 
metrically complete, so that free geodesic motion does not reach any boundary 
in finite time, {\it or} metrically incomplete but with a potential that is 
infinite at the boundary, so that no dynamical trajectory reaches it.

For a single kink in $\phi^4$ or sG theory, there is a 1-dimensional moduli 
space, with modulus the kink position (centre) $a$. The metric is 
Euclidean and the potential constant, so kink motion 
is at constant velocity. Shape mode oscillations of the $\phi^4$ kink can also 
be accommodated. Moduli space dynamics is a fundamentally non-relativistic 
approximation, so the kink is not Lorentz contracted.

The simplest moduli space for kink-antikink dynamics in $\phi^4$ theory is 
obtained using a naive superposition of the kink and antikink, with 
the centres at $-a$ and $a$ respectively. The modulus $a$ runs from $-\infty$
to $\infty$ and the moduli space is complete. An interesting
alternative moduli space uses a weighted kink-antikink superposition. 
When the weight factor is $\tanh(a)$, the configurations are identical 
to those occurring in the iterated kink scheme \cite{MOW}. This 
observation is useful, because the moduli space in terms
of $a$ is now incomplete. The reason is that the weighted configurations are 
symmetric in $a$, so the moduli space is just the half-line $a \ge 0$,
bounded by the vacuum configuration ($a=0$) where the kink and 
antikink annihilate. On the other hand, the iterated kink configuration has a 
different modulus $c = \sinh^2(a)$ that extends through $0$ down to 
$-1$. The extended moduli space in terms of $c$ is complete. 
Motion in this moduli space allows 
the kink and antikink to approach from large separation, smoothly 
pass through the vacuum configuration, stop at a configuration with 
a negative bump, and bounce back. In terms of the original position modulus 
$a$ -- just a change of coordinate -- there has been scattering from 
real to imaginary values and back, although the field remains real. 
This remarkable 90-degree scattering of the complexified position 
coordinate $a$ is reminiscent of the real 90-degree scattering 
that occurs for several types of soliton in two or 
three spatial dimensions.

We have also reconsidered the well-known moduli space that includes 
the shape modes of the $\phi^4$ kink and antikink (still preserving 
a reflection symmetry). This was introduced by Sugiyama \cite{Sug}, 
and studied by a number of 
authors, including Takyi and Weigel \cite{TW}, who clarified details of the 
metric structure but also reported a null-vector problem -- a zero in the 
metric. We have shown here that this null-vector is an artifact of the 
coordinate choice, and disappears if the shape mode amplitude $A$ is replaced 
by $B = A/\tanh(a)$ (the denominator could be any function linear in 
$a$ near $a=0$). 

In the sG case, the moduli space of naive kink-antikink superpositions 
consists, surprisingly, of the same field configurations that occur in the 
exact kink-antikink solution at the critical energy $16$ separating 
kink-antikink scattering solutions from breathers. Using the energy 
conservation law for the exact solution, we obtain a nice relation 
between the metric and potential on this moduli space (which provides 
a useful check on the calculations). The moduli space is complete, and
we have shown that the dynamics on it gives a good approximation to 
the exact scattering and breather solutions with energies above 
and below $16$. This relies on comparing the frequency vs. energy relations 
for the exact breathers and for the approximate breathers modelled by
moduli space dynamics, and comparing the exact positional shift that 
occurs in kink-antikink scattering with the moduli space approximation to this.

We have also considered kink-antikink-kink moduli spaces, where the two kinks 
are equidistant from the antikink at the origin. The moduli space 
of two naively superposed kinks and an antikink allows the kinks 
to approach and annihilate the 
antikink, leaving a single kink. However this moduli space is incomplete, and 
needs to be extended to allow the single kink to become more compressed 
(steeper). We have proposed how to do this, both in $\phi^4$ and sG 
theory. The kink position coordinates are imaginary in this extension, 
although the field again remains real. The resulting moduli spaces 
are still incomplete, having a boundary configuration that is 
a step function -- an infinitely steep kink that is only a finite 
distance from a smooth kink. However, the step function has 
infinite potential energy, so the dynamics on moduli space 
does not reach the boundary. 

In the sG case there is an exact solution having critical energy 
24 separating the kink-antikink-kink scattering solutions from 
the wobbling kink solutions. Its configurations define a useful 
alternative moduli space, but this moduli space is again incomplete 
and has to be extended to accommodate dynamics with energy larger 
than 24. This is another example where the metric and potential are simply 
related. 

We have reported preliminary tests of the 2-dimensional moduli space 
approximation to (reflection-symmetric) kink-antikink dynamics in 
$\phi^4$ theory, using our improved coordinate for the shape mode 
amplitude, but more remains to be done. The field equation, a PDE, 
has been solved numerically, and we have shown that the field 
evolution is close to a motion through the moduli space. $\phi^4$ theory is 
more complicated than sG theory, because of the shape modes and 
because of the radiation emitted. Despite much work in this area 
for more than 40 years \cite{KG}, various largely technical difficulties have 
arisen in attempts to match the field dynamics of kinks and antikinks 
to finite-dimensional moduli space models of the dynamics. In this 
paper we have clarified the geometry and full extent of various moduli 
spaces. This should lead to an improved treatment of the dynamics on 
them, and to a deeper understanding of the interesting and complicated 
phenomena observed in kink-antikink interactions.

\section*{Acknowledgements}

NSM has been partially supported by the U.K. Science and 
Technology Facilities Council, consolidated grant ST/P000681/1. 
KO acknowledges support from the NCN, Poland 
(MINIATURA 3 No. 2019/03/X/ST2/01690 and 2019/35/B/ST2/00059), 
TR and AW were supported by the Polish National Science Centre, 
grant NCN 2019/35/B/ST2/00059.

\end{document}